\documentclass[conference]{IEEEtran}

\usepackage{booktabs} % For formal tables

\usepackage[linesnumbered,ruled,vlined]{algorithm2e}
\usepackage{algpseudocode}
\usepackage{subfigure}
\usepackage{graphicx}
\usepackage{amsmath}
\usepackage{bm}
\usepackage{url}
\usepackage{stmaryrd}
\usepackage{balance}
\usepackage{amssymb}
\usepackage{pgfplots}
\usepackage{pifont}
\usepackage{epsfig}
\usepackage{booktabs}% professional tables
\usepackage{pgffor}
\usepackage{tikz}
\usepackage{multirow}
\usepackage{amsmath}
\usepackage{tikz}
\usetikzlibrary{positioning,arrows,calc}
\usepackage[all]{xy}
\usepackage{makecell}
\usepackage{enumitem}
\makeatletter
\newcommand\xleftrightarrow[2][]{%
	\ext@arrow 9999{\longleftrightarrowfill@}{#1}{#2}}
\newcommand\longleftrightarrowfill@{%
	\arrowfill@\leftarrow\relbar\rightarrow}
\makeatother

\newtheorem{defn}{\noindent\textsc{Definition}}

\newcommand{\our}{\textsc{AA-HGNN}}
\newcommand{\concatenate}{\operatornamewithlimits{\|}}

\usepackage{array}
\makeatletter
\newcommand{\thickhline}{%
	\noalign {\ifnum 0=`}\fi \hrule height 1pt
	\futurelet \reserved@a \@xhline
}
\begin{document}

%\CopyrightYear{2017} 
%\setcopyright{acmcopyright}
%\conferenceinfo{WSDM 2017,}{February 06-10, 2017, Cambridge, United Kingdom}
%\isbn{978-1-4503-4675-7/17/02}\acmPrice{\$15.00}
%\doi{http://dx.doi.org/10.1145/3018661.3018734}

%----------------------------------------------------------------------------------------
\title{Adversarial Active Learning based Heterogeneous Graph Neural Network for Fake News Detection}

\author{\IEEEauthorblockN{Yuxiang Ren\IEEEauthorrefmark{1},
		Bo Wang\IEEEauthorrefmark{2}, Jiawei Zhang\IEEEauthorrefmark{1}
		and Yi Chang\IEEEauthorrefmark{2}}
	\IEEEauthorblockA{\IEEEauthorrefmark{1}IFM Lab, Department of Computer Science, Florida State University, FL, USA\\
		\IEEEauthorrefmark{2}Artificial Intelligence, Jilin University, Jilin, China\\
		Email: yuxiang@ifmlab.org,
		bowang19@mails.jlu.edu.cn,
		jiawei@ifmlab.org,
		yichang@jlu.edu.cn}}

\maketitle
%-----------------------------------------------
\begin{abstract}
	The explosive growth of fake news along with destructive effects on politics, economy, and public safety has increased the demand for fake news detection. Fake news on social media does not exist independently in the form of an article. Many other entities, such as news creators, news subjects, and so on, exist on social media and have relationships with news articles. Different entities and relationships can be modeled as a heterogeneous information network (HIN). In this paper, we attempt to solve the fake news detection problem with the support of a news-oriented HIN. We propose a novel fake news detection framework, namely \textbf{A}dversarial \textbf{A}ctive Learning-based \textbf{H}eterogeneous \textbf{G}raph \textbf{N}eural \textbf{N}etwork (\textbf{{\our}}) which employs a novel hierarchical attention mechanism to perform node representation learning in the HIN. {\our} utilizes an active learning framework to enhance learning performance, especially when facing the paucity of labeled data. An adversarial selector will be trained to query high-value candidates for the active learning framework. When the adversarial active learning is completed, {\our} detects fake news by classifying news article nodes. Experiments with two real-world fake news datasets show that our model can outperform text-based models and other graph-based models when using less labeled data benefiting from the adversarial active learning. As a model with generalizability, {\our} also has the ability to be widely used in other node classification-related applications on heterogeneous graphs.

\end{abstract}

%\category{H.2.8}{Database Management}{Database Applications-Data Mining} 
%-----------------------------------------------

%-----------------------------------------------

\section{Introduction}\label{sec:introduction}

With the widespread use of social networks, fake news has become a serious social problem that cannot be ignored. In politics, fake news biases people's judgments about major issues like Brexit \cite{bastos2019brexit} and the 2016 US presidential election \cite{AG17}. A lot of fake news is spread on various social platforms during the 2016 US presidential election, e.g., on Facebook. 115 pro-Trump fake stories that were shared a total of 30 million times, and 41 pro-Clinton fake stories being shared a total of 7.6 million times are observed \cite{AG17}. In the economic field, the extreme sensitivity of the capital market has caused it to suffer from fake news. For instance, $\$130$ billion is wiped out in stock value after a piece of fake news claimed that then-president Barack Obama was injured in an explosion \cite{Rapoza17}. In public safety affairs, people's responses to emergencies, from natural disasters to terrorist attacks, have been disrupted by the spread of false news online \cite{mendoza2010twitter,gupta2013faking,vosoughi2018spread}. In view of this, the detection and mitigation of fake news is imperative.

However, detecting fake news on social media is particularly challenging. At first, fake news is written and published intentionally, so the content is carefully camouflaged. Fake content may account for only 1\% of news articles, but it is sufficient for the purpose. This makes it difficult to detect fake news simply based on news articles. Secondly, fake news spreads much faster than real news. According to the research in \cite{vosoughi2018spread}, many more people retweeted falsehood than they did the truth on Twitter. Therefore, the detection of fake news has high requirements for timeliness. Once a large number of users have obtained false consultations, destructive effects have already been caused. What's more, it is expensive and time-consuming to check and label the credibility of news articles by experts manually. Fake news detection methods requiring a large number of labels are not practical in the real world.

%-----------------------------------------------------------------
\begin{figure}[t]
	\centering
	\begin{minipage}[l]{1\columnwidth}
		\centering
		\includegraphics[width=\textwidth]{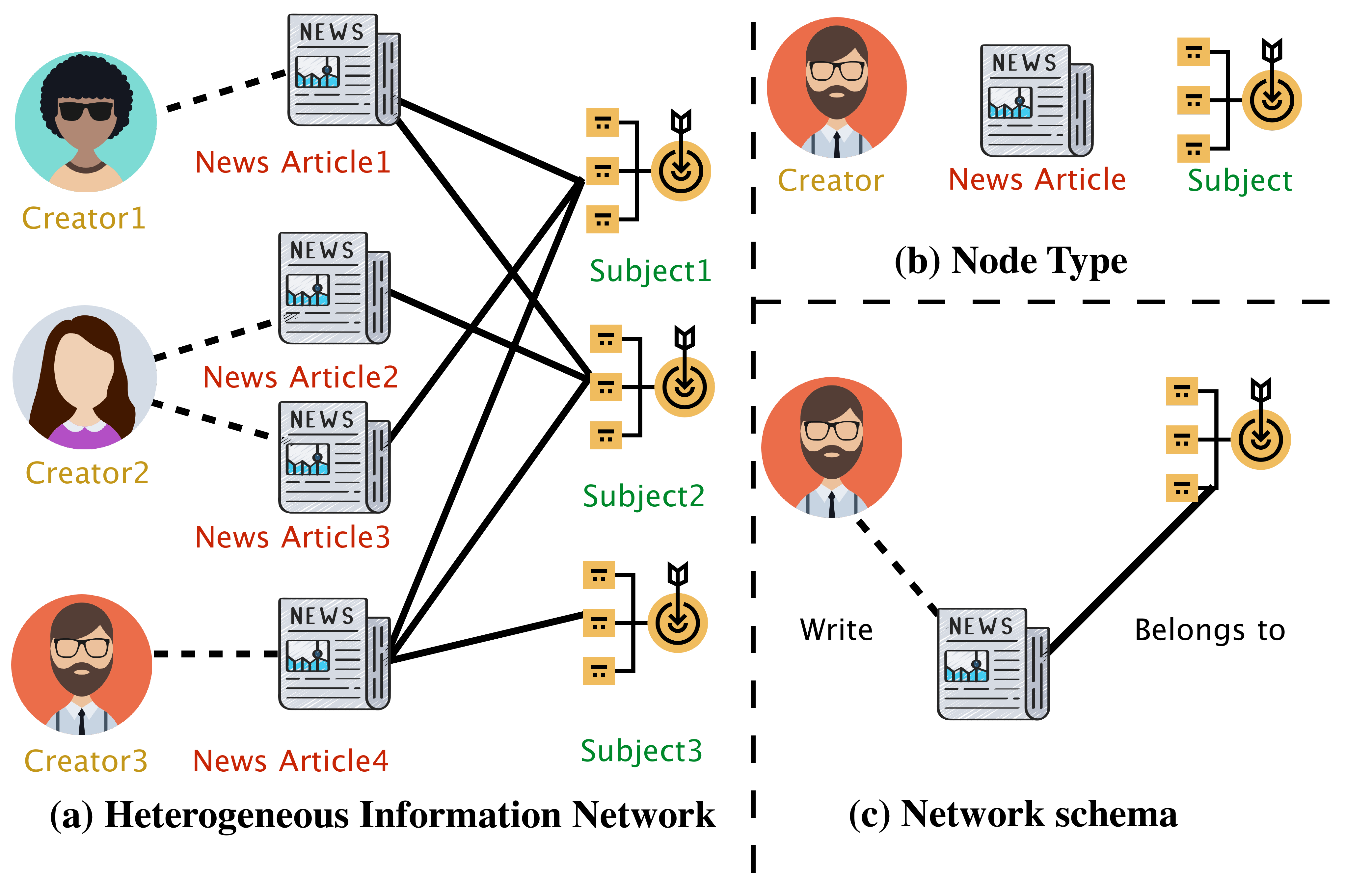}
	\end{minipage}
	\vspace{-10pt}
	\caption{An illustrative example of a heterogenous information network based on PoliticFact data (News-HIN). (a) A  News-HIN consists three types of nodes and two types of links. (b) Three types of nodes (i.e., Creator, News article, Subject). (c) Network schema of News-HIN  
	}\label{fig:example}
	\vspace{-10pt}
\end{figure}
%-----------------------------------------------------------------

On social media, focusing on news articles alone is not comprehensive, because news does not exist independently in the form of articles. In fact, there are many entities related to news articles, such as news creators, news subjects and so on. These different types of entities and their relationships provide a more comprehensive perspective on identifying news articles. A heterogeneous information network (HIN for short) \cite{SH12,SLZSY17} can be utilized to represent these entities and relationships. An illustration of such a news oriented heterogenous information network (News-HIN) based on \textit{PolitiFact}\footnote{https://www.politifact.com/} data is presented in Figure~\ref{fig:example}. In addition to the information provided in the news article, we are able to collect profile information of news creators from social networks and other supplementary knowledge libraries.
For the news subjects, the background and auxiliary knowledge can be collected to support the fake news detection. With the support of a News-HIN, fake news detection task can be formulated as the node classification problem. In this way, more sufficient information and knowledge can be used to check the credibility of news articles.
%\noindent \textbf{Problem Studied}: \underline{A}dversarial \underline{A}ctive \underline{H}IN-based \underline{F}ake \underline{N}ews \underline{D}etection 

The main challenges of the fake news detection problem in a News-HIN lie in the following points:  

\begin{itemize}[leftmargin=*]
	
	\item \textit{Paucity of Training data}: Fake news appears and spreads very quickly. The real-time nature of news also makes outdated labels worthless. Therefore, fake news detection often faces the challenge of lacking valuable training data. This requires that models can effectively detect potential fake news with the support of a small amount of training data.
	
	\item \textit{Heterogeneity}: Multiple types of heterogeneous information exist in a News-HIN, which can provide key signals for identifying fake news article nodes. At the same time, learning effective node representations in a News-HIN considering both structural and type information is non-trivial.
	
	\item \textit{Generalizability}: In order to ensure the applicability of the proposed model to diverse and possibly changing News-HINs, we need to provide a general detection model that can handle News-HINs containing any types of nodes and different schemas. 
	
\end{itemize}

\begin{figure}[t]
	\centering
	%	\vspace{-5pt}
	\begin{minipage}[l]{1.0\columnwidth}
		\centering
		\includegraphics[width=\textwidth]{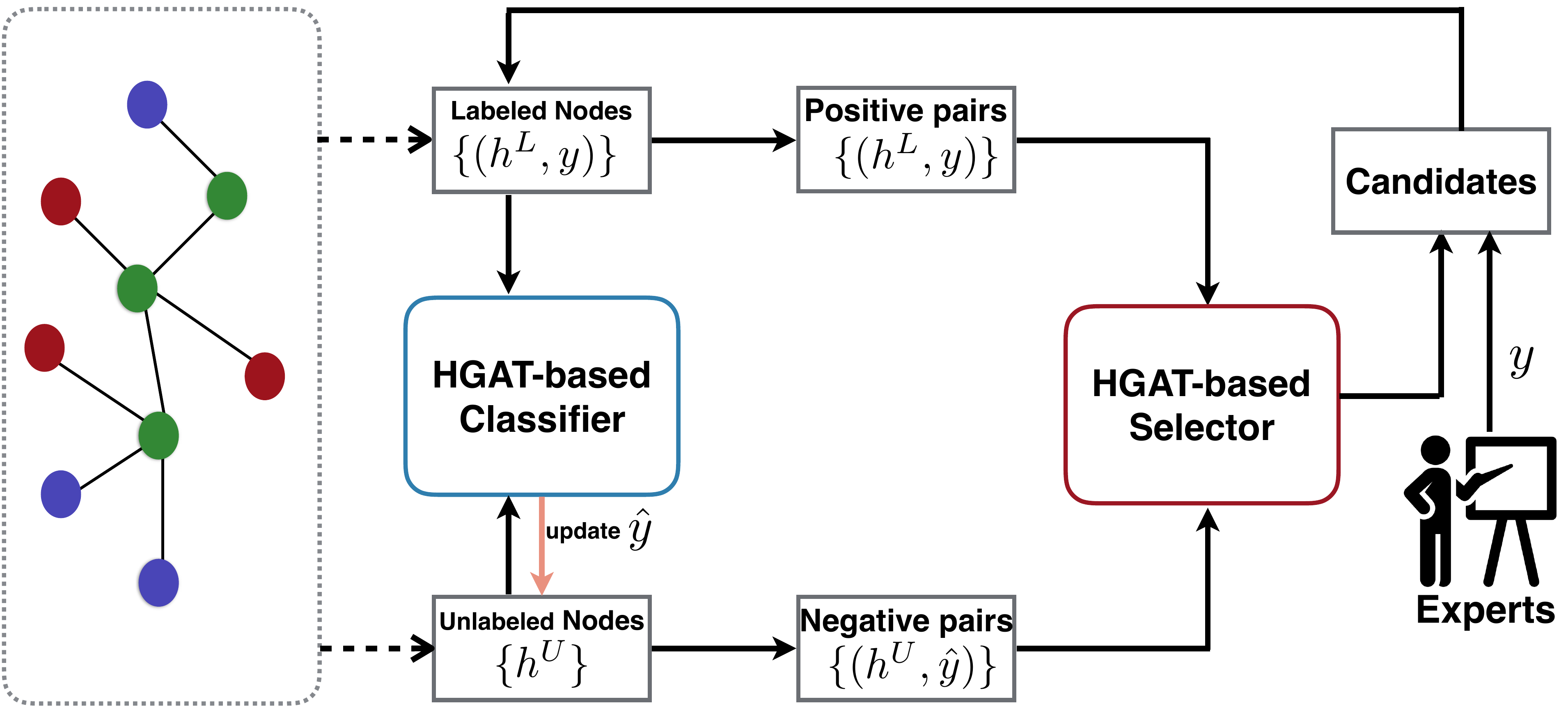}
	\end{minipage}
	\vspace{-10pt}
	\caption{Overall Framework. 
	}\label{fig:framework}
	\vspace{-18pt}
\end{figure}

To solve these challenges aforementioned, we propose a novel \textbf{A}dversarial \textbf{A}ctive Learning-based \textbf{H}eterogeneous \textbf{G}raph \textbf{N}eural \textbf{N}etwork ({\our}) to detect fake news in the News-HIN. For the first challenge, the proposed framework is built on an active learning framework, where a classifier and a selector are included. By continuously querying high-value candidate nodes for classifier training and tuning, excellent performance can be achieved with a small amount of labeled data. For the second challenge, a heterogeneous graph neural network with a novel \textbf{H}ierarchical \textbf{G}raph \textbf{A}ttention (HGAT) mechanism is utilized in both the classifier and the selector. Based on the two-level attention mechanism (node-level \& schema-level), HGAT can get the optimal combination of different types of neighbors in a hierarchical manner. The HGAT-based classifier is responsible for conducting classification on news article nodes. The HGAT-based selector is used to evaluate the predicted label from the classifier for high-value selection. The selected candidate nodes will become part of the training set via experts labeling. The classifier and the selector are trained based on adversarial learning: with the improvement of the predicted label quality by the classifier, the evaluation ability of the selector will be improved to continuously select better candidates. The overall architecture of proposed framework is shown in Figure~\ref{fig:framework}. {\our} has no limitation on the structures of News-HINs, thus it has good generalizability and can solve the third challenge well. We focus on applying {\our} to fake news detection domain in this paper, but for more general problems of node classification on heterogeneous graphs, {\our} is also applicable.
%-----------------------------------------------------------------

%-----------------------------------------------------------------
The contributions of our work are summarized as follows:
\begin{itemize}[leftmargin=*]
	\item We are the first to apply adversarial active learning to fake news detection, which can achieve excellent detection performance with much less training data. It is of great significance for fake news detection, because the urgent timeliness of fake news detection makes sufficient training data impossible.
	\item We propose a novel adversarial active learning-based framework {\our} which can handle the heterogenity of News-HINs effectively through a two-level attention mechanism. {\our} is applicable to HINs with different schemas.
	\item We conduct extensive experiments on two real-world datasets to demonstrate the effectiveness of {\our}. The results show the superiority of {\our} compared with the state-of-the-art models in detecting fake news, especially facing the paucity of training data.
\end{itemize}

%The remaining paper is organized as follows. We review the related works in Section 2. Then we introduce several important concepts and formulate the problem in Section 3. The proposed model {\our} is introduced in Section 4, whose effectiveness is evaluated in Section 5. Finally, we conclude this paper in Section 6.
%-----------------------------------------------
\vspace{-10pt}
\section{Related Work} \label{sec:related_work}
%The research topics related to this paper mainly include fake news detection and adversarial and active learning. 
%\vspace{-5pt}
\subsection{Fake News Detection}
As an emerging topic, some research works in fake news detection have been proposed. Content-based fake news detection is based primarily on the deep mining of news content. \cite{GPLJFA15,BT14} extract the knowledge, a set of (Subject, Predicate, Object) triples \cite{DGHH14}, from the news content and assess the authenticity of news by comparing them with real knowledge. However, the timeliness and integrity of the knowledge map still limit the application of them \cite{ZZ18}. Writing style is extracted and utilized to measure the credibility of news by some methods. \cite{VRTL15} employs rhetorical structure theory to evaluate the authenticity in discourse level. \cite{VBAR17,MJKJB17} capture the sentiment and readability of the news content to access the extent of falsehood. But these methods based on writing style can be hard to work in the face of carefully camouflaged fake news.

Some methods use not only the news content, but also other information related to the news. Guo et al. \cite{GCZGL18} utilize LSTM and a hierarchical attention mechanism to detect rumors, which makes use of social information through the proposed social feature. Shu et al. \cite{cui2019defend} study the explainable detection of fake news with the support of both news contents and user comments. Jin et al. evaluate news credibility within a graph optimization framework \cite{JCZL16}. Methods based on matrix factorization \cite{SWL19}, tensor factorization \cite{SRMS18}, and recurrent neural networks (RNNs) \cite{NSY18,ZCFG18, ren2020hgat} are proposed to work on the news-oriented networks. 

In this paper, we model the news content and related entities as a News-HIN. Both structural information and node content of News-HIN are utilized by {\our} to identify fake news.
\vspace{-8pt}
\subsection{Graph Neural Network}
\vspace{-2pt}
Graph Neural Networks (GNNs) learn nodes' new feature vectors through a recursive neighborhood aggregation scheme \cite{MGF05,FMAMG09,XHLJ10}. A propagation model incorporating gated recurrent units to propagate information across all nodes is proposed in \cite{YDMR16}. Recently, there is a surge of generalizing convolutional operation on the graph-structured data. Joan Bruna et al. \cite{JWAY13} extend convolution to general graphs by a novel Fourier transformation in graphs. Kipf et al. \cite{TM17} propose Graph Convolutional Network (GCN). Hamilton et al. \cite{WRJ17} introduce GraphSAGE which generates embeddings by aggregating features from a node's local neighborhood directly. Graph Attention Network (GAT) \cite{VCCRLB18} first imports the attention mechanism into graphs, which is utilized to learn the importance of neighbors and aggregates the neighbors to learn the representation of nodes in the graph. However, the above graph neural networks are presented for the homogeneous graphs. Wang et al.  \cite{WJSWCYY19, ren2019heterogeneous} consider the attention mechanism in heterogeneous graph learning through the model HAN, where information from multiple meta-path defined connections can be learned effectively. However, meta-path as a handcrafted feature limits HAN. In addition, HAN only considers different types of connections between target nodes through meta-path but ignores the use of node contents carried by different types of nodes.
\vspace{-8pt}
\subsection{Adversarial and Active Learning}
\vspace{-2pt}
The principle of adversarial learning is invented in generative adversarial networks (GANs) by Goodfellow et al. \cite{goodfellow2014generative}. Adversarial learning principle has achieved excellent performance in many different topics, such as text classification \cite{li2018learning}, information retrieval \cite{wang2017irgan}, and network embedding \cite{hu2019adversarial,dai2018adversarial}. Adversarial learning method on heterogeneous network embeddings \cite{hu2019adversarial} can be used to learn a more efficient representation of news nodes in News-HIN. However, in order to detect fake news, HeGAN \cite{hu2019adversarial} still requires a large number of labeled data to train a classifier.
Active learning is an effective way to train a model with less labeled data, because not all training samples are equally important \cite{aggarwal2014active}. The number of labels needed to learn actively can be logarithmic in the usual sample complexity of passive learning~\cite{dasgupta2005analysis} . Active learning also proves its value and robustness on different topics including recommendation systems \cite{rubens2015active}, social network alignment \cite{ren2019meta,ren2019activeiter}, image classification \cite{wang2016cost} and graph matching \cite{serratosa2015interactive}. 

In this paper, {\our} combines adversarial learning and active learning. Selectors trained in an adversarial manner can continuously select high-value candidates for active learning. The high-value candidates further improve the performance of the classifier.
%-----------------------------------------------
%\vspace{-8pt}
\section{Concept and Problem Definition} \label{sec:formulation}
%\vspace{-4pt}
%In this section, we first introduce News-HIN related terminologies and then formulate the studied problem.
%\vspace{-8pt}
\subsection{Terminology Definition}
In order to make it easier to understand related concepts, we will use the \textit{PolitiFact} data as an example to illustrate here. The \textit{PolitiFact} data contain News articles, Subjects and Creators, which can be modeled into a heterogeneous network as three types of nodes and construct different types of links based on the connections among them. We can define News Oriented Heterogeneous Information Networks (News-HIN) formally as follows:
\begin{defn}
	(News Oriented Heterogeneous Information Networks (News-HIN)): The news oriented heterogeneous information network (News-HIN) can be defined as $\mathcal{G} = (\mathcal{V}, \mathcal{E})$, where the node set $\mathcal{V} = \mathcal{C} \cup \mathcal{N} \cup  \mathcal{S}$. \ \  $\mathcal{C}, \mathcal{N}$ and $\mathcal{S}$ represent Creators, News articles and Subjects respectively. We will define different types of nodes in detail later. The link set $\mathcal{E} = \mathcal{E}_{c,n} \cup \mathcal{E}_{n,s}$ involves the "Write" links between creators and news articles, and the "Belongs to" links between news articles and subjects.
\end{defn}

News articles refer to the news content post on social media or public platforms. We can define news articles in a formal way as:

\begin{defn}
	(News Articles): The News articles set can be represented as $\mathcal{N} = \{n_1, n_2,\cdots, n_m\}$. For each news article $n_i \in \mathcal{N}$, it contains its textual contents.
\end{defn}

The credibility label of $n_i$ takes value from the label set $\mathcal{Y} = \{Fake, Real\}$. In this paper, the original 
label set contains 6 different class labels (True, Mostly True, Half True, Mostly False, False, Pants on Fire). We group the labels {Pants on Fire, False, Mostly False} as fake news and group {True, Mostly True, Half True} as real news. 
Subjects denote the central ideas of news articles, which normally are the main objectives of writing news articles.

\begin{defn}
	(Subjects): The set of subjects can be denoted as $\mathcal{S} = \{s_1, s_2,\cdots, s_k\}$. For each subject $s_i \in \mathcal{S}$, it contains the textual description.
\end{defn}

Creators denote people who write news articles. We can also define this concept in a formal way.

\begin{defn}
	(Creators): The set of creators can be represented as $\mathcal{C} = \{c_1, c_2,\cdots, c_n\}$. For each creator $c_i \in \mathcal{C}$, it contains the profile information.
\end{defn}
In the PolitiFact dataset, the creators have the profile containing their titles, political party membership, and geographical residential locations. The profile information can be described by a sequence of words. 

In order to better understand the News-HIN and utilize type information, it is necessary to define the schema-level description. The schema of News-HIN serves for learning the importance of nodes and links with different types.

\begin{defn}\label{def:schema}(News-HIN Schema): 
	Formally, the schema of the given News-HIN $\mathcal{G} = (\mathcal{V}, \mathcal{E})$ can be represented as $S_{\mathcal{G}} = (\mathcal{V}_{T}, \mathcal{E}_{T})$, where $\mathcal{V}_{T}$ and $\mathcal{E}_{T}$ denote the set of node types and link types in the network respectively. Here, $\mathcal{V}_{T} = \{\phi_n,\phi_c,\phi_s\}$ and $\mathcal{E}_{T}= \{\textit{Write}, \textit{Belongs to}\}$.
\end{defn}

An illustration of News-HIN Schema based on the PolitiFact data is shown in Figure~\ref{fig:example}(c).

\subsection{Problem Formulation}\label{subsec:problem_definition}
Given a News-HIN $\mathcal{G} = (\mathcal{V}, \mathcal{E})$, the fake news detection problem aims at learning a classification function $\mathit{f}:\mathcal{N}\rightarrow\mathcal{Y}$ to classify news article nodes in the set $\mathcal{N}$ into the correct class with the credibility label in $\mathcal{Y}$. 
%Various kinds of heterogeneous information in the News-HIN $\mathcal{G}$ should be effectively incorporated, including both the textual information and network structure information. 
The news article nodes with labels can be grouped as a labeled set $\mathcal{L}$ and the rest news article nodes will be denoted as the unlabeled set $\mathcal{U} = \mathcal{N} \setminus \mathcal{L}$. Based on the active learning setting, we are also allowed to query for labels of news article nodes in $\mathcal{U}$ with a upper limit budget $b$. We also want to propose a mechanism to achieve an optimal query set $\mathcal{U}_q$ to improve the classification function $\mathit{f}:\mathcal{N}\rightarrow\mathcal{Y}$.
To resolve the above fake news detection problem, we will introduce the proposed adversarial active learning based heterogeneous graph neural network {\our} in Section~\ref{sec:method}.
%-----------------------------------------------
%-----------------------------------------------------------------
%\begin{figure*}[t]
%	\centering
%	\begin{minipage}[l]{1.8\columnwidth}
%		\centering
%		\includegraphics[width=\textwidth]{./figures/method/Adv_Structure.pdf}
%	\end{minipage}
%	\caption{The overall framework of {\our}. (a) All types of nodes are projected into a unified feature space and the weights of node pairs can be learned via node-level attention. (b) Joint learning the weight of each type of schema nodes and fuse the node representation via schema-level attention. (c) Calculate the loss and end-to-end optimization based on the classification result.}\label{fig:structure}
%\end{figure*}

%\vspace{-8pt}
\section{Proposed Method}\label{sec:method}
%\vspace{-4pt}
In this section, we propose a novel \textbf{A}dversarial \textbf{A}ctive Learning based \textbf{H}eterogeneous \textbf{G}raph \textbf{N}eural \textbf{N}etwork ({\our}) to detect fake news. As shown in Figure~\ref{fig:framework}, {\our} consists of two major components: (1) HGAT-based classifier, and (2) HGAT-based selector. We begin with the overview of the model, followed by detailed descriptions of the hierarchical graph attention neural network (HGAT). Then we illustrate the HGAT-based classifier and HGAT-based selector respectively. At last, we elaborate on the optimization of {\our}.  

\vspace{-5pt}
\subsection{Model Overview}\label{subsec:overall_framework}
The architecture of {\our} is shown in Figure~\ref{fig:framework}. The News-HIN $\mathcal{G}$ is the input of the HGAT-based classifier. $h^L$ and $h^U$ denote the initial feature of a labeled node and an unlabeled node respectively. The HGAT-based classifier is trained with both labeled and unlabeled data to predict labels $\{\hat{y}\}$ for unlabeled news article nodes. The HGAT-based selector evaluates the quality of predicted labels and selects high-value candidates from them based on a query strategy. We take the pairs of labeled nodes and their ground-truth labels $\{y\}$ as positive samples, and the pairs of unlabeled nodes and their predicted labels $\{\hat{y}\}$ are used as negative samples. A portion of positive and negative pairs are sampled to train the HGAT-based selector. After being trained, the selector outputs the confidence $\mathcal{P}$ of pairs in the test set. Based on the confidence, the proposed selection strategy selects a set of high-value unlabeled nodes as candidates with the size $k$. These candidates will be labeled by experts. In our experiments, these candidates will be moved to the training set before next round optimization. A query budget $b$ is pre-specified for {\our}. When the query budget $b$ is exceeded, the adversarial active learning stops. 

Since Hierarchical Graph Attention Neural Network (HGAT) is the basis of the classifier and the selector, which is the key to handling the heterogeneity, we will first introduce HGAT in detail in the next section.

\vspace{-5pt}
\subsection{Hierarchical Graph Attention Neural Network (HGAT)}\label{subsec:hgat}
\begin{figure}[t]
	\vspace{-10pt}
	\centering
	\begin{minipage}[l]{0.9\columnwidth}
		\centering
		\includegraphics[width=\textwidth]{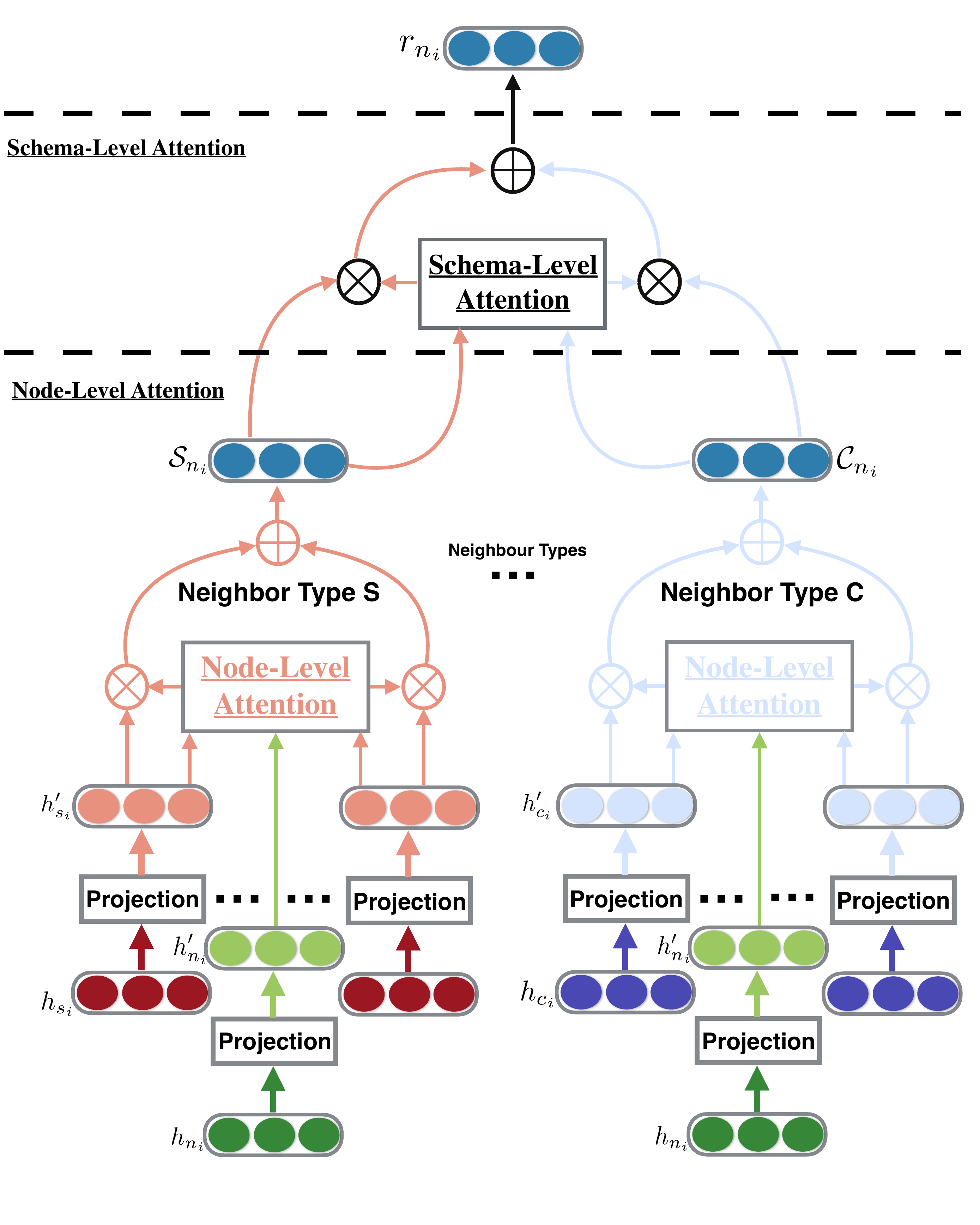}
	\end{minipage}
	\vspace{-20pt}
	\caption{Hierarchical Graph Attention Neural Network. 
	}\label{fig:hgat}
	\vspace{-20pt}
\end{figure}

The novel HGAT employs a two-level attention mechanism including node-level attention and schema-level attention. The structure of HGAT is shown in Figure~\ref{fig:hgat}. Node-level attention is responsible for learning the weights of neighbors belong to the same type and aggregates them to get the type-specific neighbor representation. Schema-level attention enables HGAT to learn the information of node types and get the optimal weighted combination of the type-specific neighbor representations. Through the two-level attention mechanism, the representations of news article nodes contain both the structural and node content information.
%-----------------------------------------------------------------
%\vspace{-30pt}

%\vspace{-10pt}
\subsubsection{Node-level attention}\label{subsec:nodeattention}
The node-level attention can learn the importance of neighbors belong to the same type respectively for each news article node $n_i \in \mathcal{N}$, and then aggregates the representation of same-type neighbors to form an integrated representation which we define as a schema node. 

%For the initial feature vectors of nodes with different types, they belongs to feature spaces with different dimensions. 
The inputs of the node-level attention layer are the node initial feature vectors $\{h\}$. Because multiple types of nodes exist in the News-HIN, the initial feature vectors belong to feature spaces with different dimensions. In order to enable the attention mechanism to output comparable and meaningful weights between different types of nodes, we first utilize a type-specific transformation matrix to project features with different dimensions into the same feature space. We take the news article node $n_i \in \mathcal{N}$ as an example. The transformation matrix for type $\phi_n$ is $\mathbf{M}^{\phi_n} \in \mathbb{R}^{F \times F^{\phi_n}}$, where $F^{\phi_n}$ is the dimension of the initial feature $h_{n_i} \in \mathbb{R}^{F^{\phi_n}}$ of the news article node $n_i$ and $F$ is the dimension of the feature space mapped to. The projection process can be shown as follows:
\begin{equation}
\begin{aligned}
h'_{n_i} &= \mathbf{M}^{\phi_n} \cdot h_{n_i}
%h'_{c_i} &= \mathbf{M}^{\phi_c} \cdot h_{c_i} \\
%h'_{s_i} &= \mathbf{M}^{\phi_s} \cdot h_{s_i}
\end{aligned}
\end{equation}

The $h'_{n_i}$ is the projected feature of node $n_i$. The $F$ is the same for all type-specific transformation matrices. Through the type-specific projection operation, the feature space of nodes with different types can be unified where the self-attention mechanism can work on to learn the weight among various kinds of nodes. 

%Here, the node-level attention will learn the importance of each type of neighbor nodes respectively. 
In the face of fake news detection, the target node is the news article node $n_i \in \mathcal{N}$. The neighbors of it belong to $\mathcal{N}\cup\mathcal{S}\cup\mathcal{C}$. It should be noted that we also regard the target node itself as a neighbor node to cooperate the self-attention mechanism. We let $T \in \{\mathcal{N}, \mathcal{S}, \mathcal{C}\}$ and nodes in $T$ have the type $\phi_t$. For $n_i$'s neighbor nodes in $T$, the node-level attention can learn the importance $\mathit{e}^{\phi_t}_{ij}$ which means how important node $t_j \in T$ will be for $n_i$. The importance of the node pair $(n_i, t_j)$ can be formulated as follows:

\begin{equation}
\mathit{e}^{\phi_t}_{ij} = \mathit{att} (h'_{n_i},h'_{t_j};\phi_t)
\end{equation}

Here, the node-level attention $\mathit{att}$ denotes the same deep neural network as \cite{VCCRLB18}. $\mathit{att}$ is shared for all neighbor nodes with the same type $\phi_t$. The masked attention captures the network structure information where only node $t_j \in neighbor_{n_i}$ (being neighbors of node $n_i$) will be calculated and recorded as $\mathit{e}^{\phi_t}_{ij}$. Otherwise, the attention weight will be 0. We normalize them to get the weight coefficient $\alpha^{\phi_t}_{ij}$ via softmax function:
%-----------------------------------------------------------------

\begin{figure}[t]
	\vspace{-10pt}
	\centering
	\begin{minipage}[l]{1\columnwidth}
		\centering
		\includegraphics[height=0.5\textwidth]{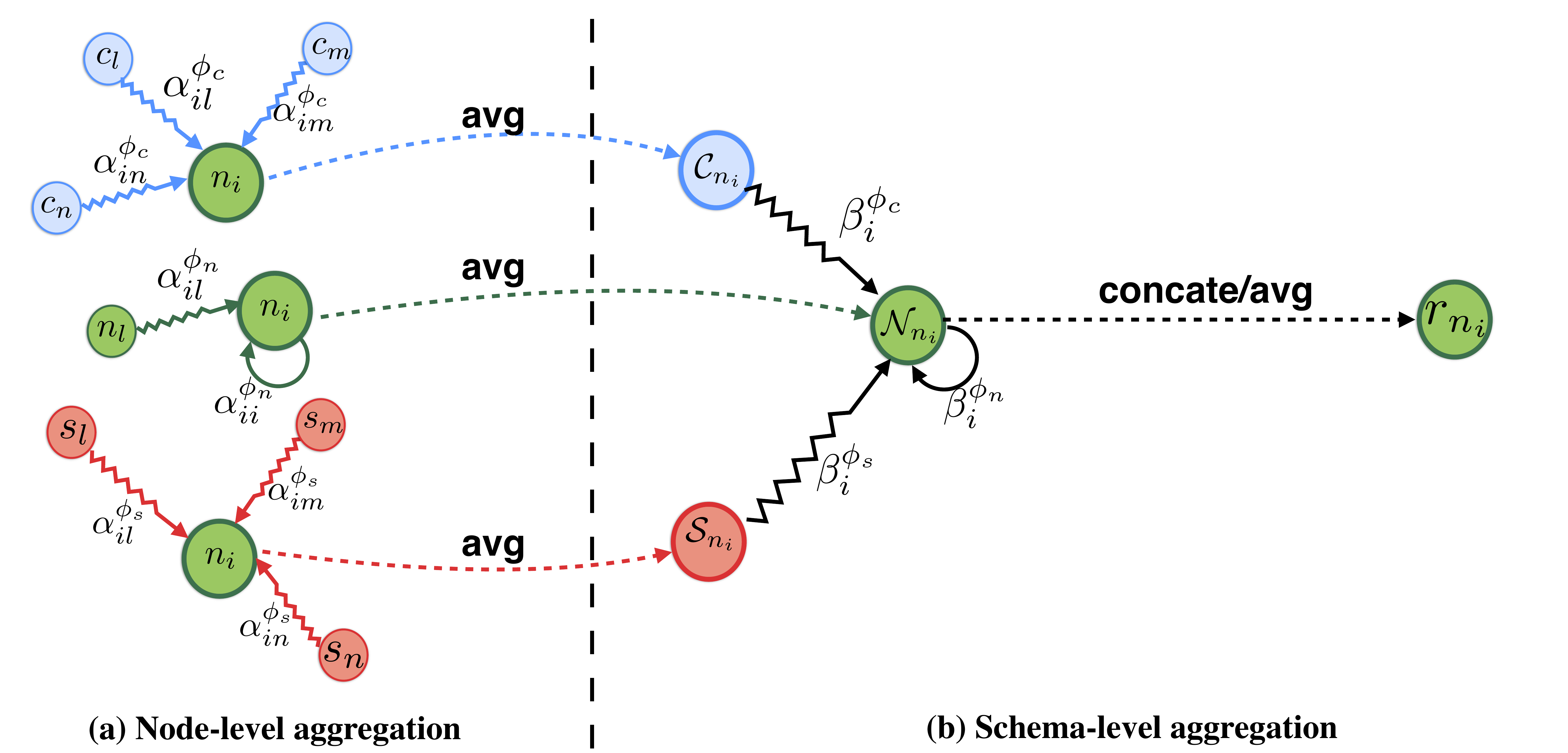}
	\end{minipage}
	\vspace{-10pt}
	\caption{Explanation of aggregating process in node-level and schema-level.}\label{fig:aggregation}
	\vspace{-15pt}
\end{figure}
%-----------------------------------------------------------------
\begin{equation}
\alpha^{\phi_t}_{ij} = \textup{softmax}_j(\mathit{e}^{\phi_t}_{ij}) = \frac{exp(\mathit{e}^{\phi_t}_{ij})}{\sum_{t_k \in neighbor_{n_i}}\mathit{e}^{\phi_t}_{ik}}
\end{equation}

Then, the schema node $T_{n_i}$ can be aggregated by the neighbor's projected features with the corresponding weights as follows:
\begin{equation}
T_{n_i} = \sigma(\sum_{t_j \in neighbor_{n_i}}\alpha^{\phi_t}_{ij} \cdot h'_{t_j})
\end{equation}

Similar to Graph Attention Network (GAT) \cite{VCCRLB18}, a multi-head attention mechanism can be used to stabilize the learning process of self-attention in node-level attention. In details, $K$ independent node-level attentions execute the transformation of Equation (4), and then the features achieved by $K$ heads will be concatenated, resulting in the output representation of the schema node:
\begin{equation}
T_{n_i} = \concatenate_{k=1}^K \sigma(\sum_{t_j \in neighbor_{n_i}}\alpha^{\phi_t}_{ij} \cdot h'_{t_j})
\end{equation}
where $\concatenate$ represents concatenation. In the problem we face, every target node $n_i$ has 3 schema nodes corresponding to 3 different types neighbors (include itself) based on the Definition~\ref{def:schema}. They can be denoted as $\mathcal{N}_{n_i}$, $\mathcal{C}_{n_i}$, $\mathcal{S}_{n_i}$.
\subsubsection{Schema-level attention}\label{subsec:schemaattention}
Through the node-level attention, we fuse information from neighbor nodes with the same type into the representation of a schema node. Now, HGAT needs to learn the representation of news article nodes from all schema nodes. Different schema nodes contain type-specific information, which requires us to learn the importance of different node types. Here, the schema-level attention is proposed to learn the importance of different schema nodes, and finally use the learned coefficients for weighted combination.

In order to obtain sufficient expressive power to calculate the attention weights between schema nodes, one learnable linear transformation is applied to the schema nodes. The linear transformation is parametrized by a weight matrix $\mathbf{W} \in  \mathbb{R}^{F' \times KF}$, where $K$ is the number of heads in node-level attention. The schema-level attention $\mathit{schema}$ is a single-layer feedforward neural network applying the activating function Sigmoid with the dimension $2F'$. For the schema node $T_{n_i}$, the importance of it can be denoted as $\mathit{w}^{\phi_t}_{i}$:

\begin{equation}
\mathit{w}^{\phi_t}_{i} = \mathit{schema} (\mathbf{W}T_{n_i},\mathbf{W}\mathcal{N}_{n_i})
\end{equation}

We normalize the imoportance of each schema nodes through a softmax function. Then coefficients of the final fusion can be denoted as $\beta^{\phi_t}_{i}$, which can be calculated as follows:

\begin{equation}
\beta^{\phi_t}_{i} = \textup{softmax}_t(\mathit{w}^{\phi_t}_{i}) = \frac{\textup{exp}(\mathit{w}^{\phi_t}_{i})}{\sum_{\phi \in \mathcal{V}_{T}}\textup{exp}(\mathit{w}^{\phi}_{i})}
\end{equation}

Based on the learned coefficients, we can fuse all schema nodes to get the final representation $r_{n_i} \in  \mathbb{R}^{F'}$ of the target node $n_i$:
\begin{equation}
r_{n_i} = \sum_{\phi_t \in \mathcal{V}_{T}}\beta^{\phi_t}_{i} \cdot T_{n_i}
\end{equation}

%The set of learned final representation is denoted as $\mathcal{R}$. 
We also describe the two-level aggregating process in Figure~\ref{fig:aggregation} for reference.

\begin{figure}[t]
	\vspace{-10pt}
	\centering
	\begin{minipage}[l]{1\columnwidth}
		\centering
		\includegraphics[width=\textwidth]{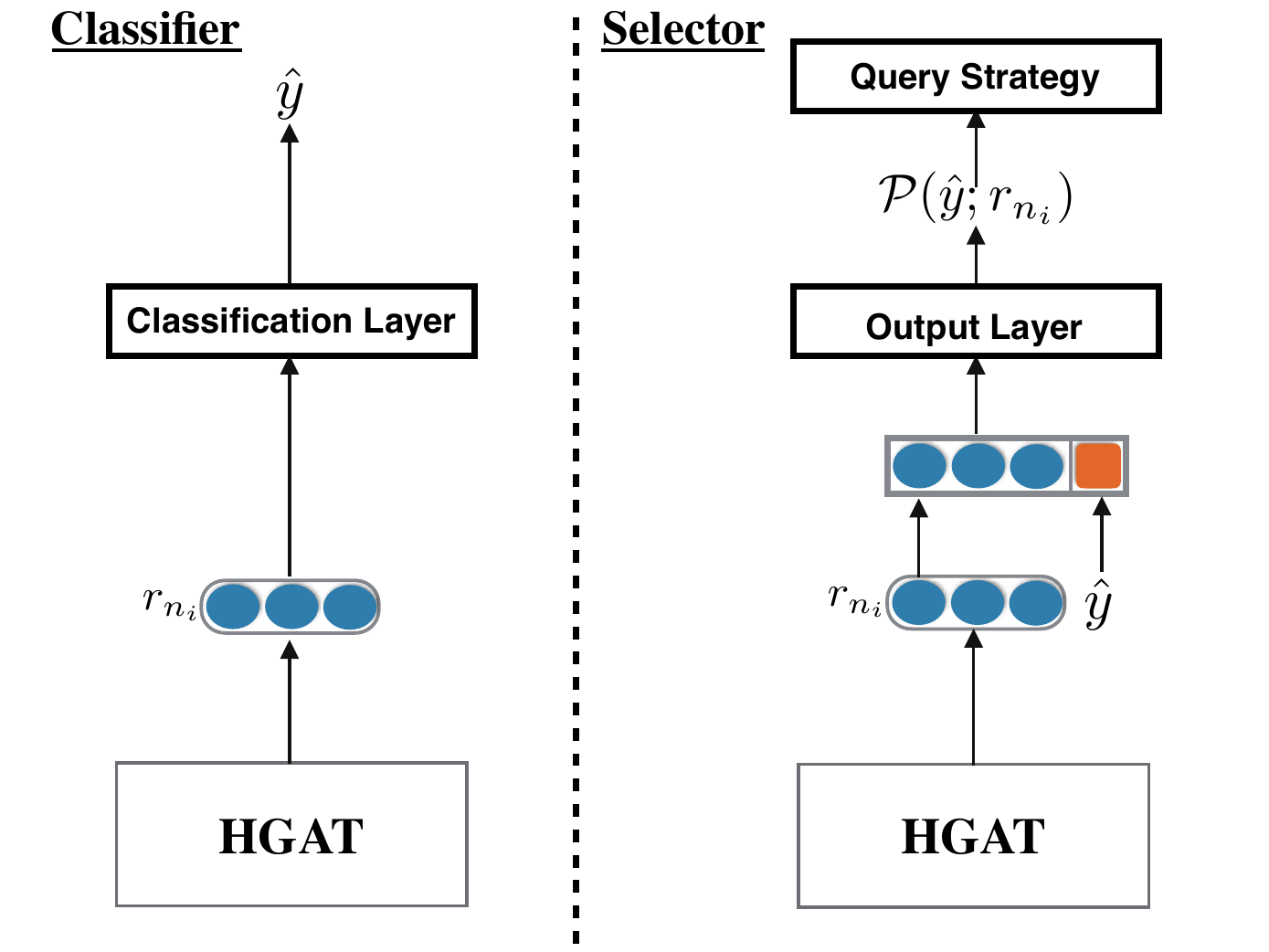}
	\end{minipage}
	\vspace{-10pt}
	\caption{HGAT-based Classifier and HGAT-based Selector. 
	}\label{fig:cla_sel}
	\vspace{-20pt}
\end{figure}
\subsection{HGAT-based Classifier}\label{subsec:classifier}
As shown in the left side of Figure~\ref{fig:cla_sel}, HGAT and a classification layer constitute a HGAT-based classifier. The input of HGAT-based classifier is the same as HGAT, which are the initial feature vectors of nodes. The classification layer can output the predicted labels $\{\hat{y}\}$ of unlabeled news article nodes. In our experiments, a logistic regression layer works as the classification layer. 

For the fake news detection tasks, the optimization objective function of the HGAT-based classifier can leverage the cross-entropy loss minimization. The HGAT-based classifier can be optimized in an end-to-end manner by backpropagation. We define the set of labeled news article nodes as $\mathcal{N}_L$ and the set of unlabelled news article nodes as $\mathcal{N}_U$, then the cross-entropy loss we used can be written as:

\begin{equation}
%\mathit{Loss}(\mathcal{R}, \mathcal{N}_L) 
\mathit{Loss}_{classifier} = - \sum_{n_i \in \mathcal{N}_L} (y_{n_i} \textup{log}(p_{n_i}) + (1-y_{n_i})\textup{log}(1-p_{n_i}))
\label{eq:loss}
\end{equation} 
Here, $y_{n_i}$ is a binary indicator (0 or 1) indicating if the binary class label is the correct classification for the news article node representation $r_{n_i}$. $p_{n_i}$ is the predicted probability of labeled news article node $n_i$.

When the optimization is completed, the predicted probability of unlabeled news article nodes in $\mathcal{N}_U$ are rounded and cast into predicted labels $\{\hat{y}\}$. The predicted labels $\{\hat{y}\}$ will be evaluated by the HGAT-based selector which is described in the next section.  
\subsection{HGAT-based Selector}\label{subsec:selector}
The structure of a HGAT-based selector is shown in the right side of Figure~\ref{fig:cla_sel}. The inputs of the layers of HGAT are the initial feature vectors $\{h\}$. Based on the learned representation $r_{n_i}$, we then concatenate $r_{n_i}$ with the predicted label $\hat{y}$ (or the ground-truth label $y$ of the labeled node). We denote this concatenated vector as $z_{n_i} \in  \mathbb{R}^{(F'+1)}$:
\begin{equation}
z_{n_i} = \left[r_{n_i},\hat{y}\right]
\label{eq:conc}
\end{equation}
\vspace{-12pt}
\begin{algorithm}[t]
	
	\caption{Adversarial Active optimization of {\our}}
	%	\large
	\small 
	\label{alg:aa_opti}
	\KwIn{The News-HIN $\mathcal{G} = (\mathcal{V}, \mathcal{E})$; The set of labeled news article nodes $\mathcal{N}_L$; The set of unlabeled news article nodes $\mathcal{N}_U$; The query budget $b$; The query batch size $k$; Number of samples $m$;		
	}
	%	\KwOut{The set of predicted labels $\{\hat{y}\}$}
	%	\Begin{
	$\mathcal{U}_q$ = $\emptyset$\;		
	\While{$|\mathcal{U}_q| < b$}{
		\Comment{\textbf{Optimization for HGAT-based classifier}}\; 
		\Begin{
			Train the HGAT-based classifier on $\mathcal{N}_L$ via Eq.\ref{eq:loss}\;
			Predict the labels of nodes in $\mathcal{N}_U$\;
			Update the set of predicted labels $\{\hat{y}\}$\;
		}
		\Comment{\textbf{Optimization for HGAT-based selector}}\; 
		\Begin{ 
			Sample $m$ nodes from $\mathcal{N}_L$ to construct positive samples via Eq.\ref{eq:conc}, i.e., $z_{n_j}, n_j \in \mathcal{N}_L$\;
			Sample $m$ nodes from $\mathcal{N}_U$ to construct negative samples via Eq.\ref{eq:conc}, i.e., $z_{n_k}, n_k \in \mathcal{N}_U$\;
			Train the HGAT-based selector on positive and negative samples\;
			Predict the probability $\mathcal{P}$ via Eq.\ref{eq:loss_sel}\;
			Query $k$ candidates based on Definition \ref{def:query}\;
		}
		$\mathcal{U}_q$ = $\mathcal{U}_q \cup \{candidates\}$\;
		Labeling $k$ candidates by experts\;
		$\mathcal{N}_L$ = $\mathcal{N}_L \cup \{candidates\}$\;	
		$\mathcal{N}_U$ = $\mathcal{N}_U \setminus \{candidates\}$\;	
	}
	\Return{The set of predicted labels $\{\hat{y}\}$}
	%	}
	
\end{algorithm}

The purpose of the HGAT-based selector is to evaluate the probability that how likely the $z_{n_i}$ is from the set of labeled news article nodes $\mathcal{N}_L$. A higher possibility represents that a news article node matches the predicted label better. At the same time, if a node does not match the predicted label, it is likely to indicate that the predicted label is wrong. The output layer is responsible for predicting the probability $\mathcal{P}(\hat{y};r_{n_i})$. Here, we use a logistic regression layer as the output layer. We sample $z_{n_j}, n_j \in \mathcal{N}_L$ as the positive samples, and the same number of $z_{n_k}, n_k \in \mathcal{N}_U$ are sampled as the negative samples. These positive and negative samples constitute the training set for the HGAT-based selector. The loss function used by HGAT-based selector is a cross-entropy loss:

\begin{equation}
\mathit{Loss}_{selector} = - \sum (y \textup{log}(\mathcal{P}) + (1-y)\textup{log}(1-\mathcal{P}))
\label{eq:loss_sel}
\end{equation} 
$y \in \{0, 1\}$ denotes the negative-positive label of the concatenated vector in training set. $\mathcal{P}$ is the predicted probability of label being positive. This loss function can be optimized by backpropagation.

%The rest concatenated vectors of unlabeled news article nodes are in the testing set.  
%%-----------------------------------------------------------------
%\begin{figure}[h]
%	\vspace{-10pt}
%	\centering
%	\begin{minipage}[l]{0.9\columnwidth}
%		\centering
%		\includegraphics[width=\textwidth]{./figures/method/HGAT_sel.pdf}
%	\end{minipage}
%	\vspace{-10pt}
%	\caption{HGAT-based Selector. 
%	}\label{fig:selector}
%	\vspace{-10pt}
%\end{figure}

The rest concatenated vectors of unlabeled news article nodes are in the testing set. After training, the HGAT-based selector will output the probability $\mathcal{P}$ for testing samples. 

Based on the probability, we propose a query strategy to select high-value candidates for active learning. As we mentioned before, a lower probability $\mathcal{P}$ indicates that the unlabeled news article node and the predicted label do not match. It also represents there is a high probability that the predicted label will be wrong. Obviously, if the news article node we query was not able to be classified correctly by the HGAT-based classifier, then it will be more "informative" than the nodes that have been correctly classified. Besides, we can make it as part of the training set in the next round of training after experts labeling, thereby correcting the misclassified nodes in the test set for similar reasons. So the query strategy is: 
\begin{defn}(Query Strategy):
	All samples in the test set will be sorted in ascending order according to the predicted probability $\mathcal{P}$, the top $k$ candidates will be added to $\mathcal{U}_q$. Here, $k$ denotes the query batch size. 
	\label{def:query}
\end{defn}
%\vspace{-10pt}
\subsection{Adversarial Active Optimization}\label{subsec:opt}
In {\our}, the HGAT-based classifier and the HGAT-based selector cooperate in an adversarial active manner. We adopt the iterative optimization to train these components in {\our}. In each iteration, the HGAT-based classifier and the HGAT-based selector have trained alternately. Specifically, we first train the HGAT-based classifier to output the predicted labels. Then the HGAT-based selector will be trained by the predicted labels from the classifier. Based on the optimized selector, $k$ candidates will be queried in one iteration and be added to $\mathcal{U}_q$ used as training data in the next iteration. Each time $k$ candidates are obtained, the classification performance of the HGAT-based classifier can be improved in the next iteration. As a consequence, the credibility of predicted labels will be increased. Better predicted labels further improve the evaluation performance of the HGAT-based selector. We repeat the above iteration until the size of $\mathcal{U}_q$ exceeds the query budget $b$. The adversarial active optimization of {\our} is described in Algorithm~\ref{alg:aa_opti}.

%-----------------------------------------------
%\vspace{-20pt}
\section{Experiments}\label{sec:experiment}
%\vspace{-4pt}
To test the effectiveness of {\our}, extensive experiments are designed and conducted on two real-world fake news datasets. We first introduce the datasets. Then experimental settings are provided. We aim
to answer the following evaluation questions based on experimental results together with the detailed analysis:
\begin{itemize}[leftmargin=*]
	\item \textbf{Question 1}: Can {\our} improve fake news detection performance by modeling data as a News-HIN? 
	\item \textbf{Question 2}: Can Hierarchical Graph Attention (HGAT) mechanism handle the heterogeneity of the News-HIN effectively?
	\item \textbf{Question 3}: Can the active learning setting of {\our} overcome the paucity of training data?
	\item \textbf{Question 4}: Can adversarial learning between the classifier and the selector significantly help improve the performance?
\end{itemize}

\subsection{Dataset Description}
%\vspace{-4pt}
%-----------------------------------------------------------------------
% Table: Dataset
%\vspace{-10pt}
\begin{table}[h]
	\vspace{-10pt}
	\caption{Properties of the Heterogeneous Networks}
	\vspace{-5pt}
	\label{tab:datastat}
	\centering
	\scriptsize
	\begin{tabular}{cll|ll}
		%------------------------------------------
		\toprule	
		&\multicolumn{2}{c}{PolitiFact Network}  &\multicolumn{2}{c}{BuzzFeed Network}\\
		\midrule 
		%------------------------------------------
		\multirow{3}{*}{\# node}
		&article   & 14,055 &article   & 182  \\
		&creator  & 3,634 &twitter user   & 15,257 \\
		&subject & 152 &publisher   & 9 \\
		\midrule 
		\multirow{2}{*}{\# link}
		&creator-article    &14,055  &publisher-article    &182   \\
		&article-subject    & 48,756 &article-twitter user    & 25,240 \\
		%------------------------------------------
		\bottomrule
	\end{tabular}
	\vspace{-10pt}
\end{table}
We use two datasets to verify our model in experiments. The main dataset is collected from the platform with fact-checking: \textit{PolitiFact}, which is operated by Tampa Bay Times. The news after fact-checking from \textit{PolitiFact} mainly are the statements or news articles posted by the politicians (Congress members, White House staffs, lobbyists) and political groups. They are creators of news articles in our experiments. Regarding these news articles, \textit{PolitiFact} will provide the original contents, fact-checking results and comprehensive fact-checking reports on the website. When presenting these news articles, the platform will categorize them into different subjects based on contents and topics. A brief description of each subject will be provided as well. The fact-checking results can indicate the credibility of corresponding news articles and take values from \{True, Mostly True, Half True, Mostly  False, False, Pants on Fire!\}. In the \textit{PolitiFact} dataset, $1322$ news articles are marked as "Pants on Fire", while the number of news articles with "False" is $2601$. Besides, $2539$ "Mostly False" news articles and $2765$ "Half True" news articles exist in the dataset. The number of "Mostly True" and "True" news is $2676$ and $2149$ respectively. We group the labels \{Pants on fire, False, Mostly False\} as fake news and group \{True, Mostly True, Half True\} as real news, the quantity of fake news is $6465$ corresponding to $7590$ real news. The fact-checking results will be used as the ground truth in experiments. We won't make use of comprehensive fact-checking reports in this paper. 
We have established a heterogeneous information network based on the \textit{PolitiFact} dataset. The HIN includes three types of nodes: article, creator and subject and two types of links: Write (between article and creator) and Belongs to (between article and subject). In order to verify the generalization and stability of {\our}, we use a public dataset \textit{BuzzFeed}\footnote{https://github.com/KaiDMML/FakeNewsNet/tree/old-version
} from Shu et al.\cite{shu2019beyond}. \textit{BuzzFeed} contains $91$ real news articles and $91$ fake news articles. We also construct a HIN based on \textit{BuzzFeed} dataset. There exist three types of nodes: article, twitter user and publisher. The key statistical data describing the HINs can be found in Table~\ref{tab:datastat}.

%-----------------------------------------------------------------------
\vspace{-5pt}
\subsection{Experimental Settings}

\subsubsection{Experimental Setup}
In the experiments, we are able to acquire the set of news article nodes which are the target node to conduct the classification. For the \textit{PolitiFact} dataset, the fact-checking results corresponding to news articles are used as the ground truth for model learning and evaluation. We group fact-checking results \{Pants on fire, False, Mostly False\} as a Fake class and group \{True, Mostly True, Half True\} as a Real class. Because our target is to detect fake news, we treat Fake class as the positive class and Real class as the negative class. For all comparison methods, we use 20\% of news article nodes as the training set and 10\% of the nodes as the validation set. In addition, the testing ratio is fixed as 10\%. For {\our}, we use 1000 nodes to initialize the active learning. The query budget $b$ is 1800 and the query batch size $k$ is 200. In this way, 2800 nodes (20\% of news article nodes) are utilized to train {\our} finally. 
\textit{BuzzFeed} dataset has only two types of labels: True and fake, we can use it directly. The rest setting is the same as the \textit{PolitiFact} dataset. We run the experiments on a Dell PowerEdge T630 Server with 2 20-core Intel CPUs and 256GB memory and the other Server with 3 GTX-1080 ti GPUs. Code is available at the \textit{link}\footnote{https://www.dropbox.com/sh/bmgz7d1q3tq5429/AAAAcmbgKOp-gtftVWhz533ua?dl=0}.

\subsubsection{Data Preprocessing}

Two datasets both contain textual data with different length. In order to fit to the non-sequential models, we have to transform the input features of each type of nodes
into a vector with a fixed length. To deal with the problem, we use \textit{TfidfVectorizer} in \textit{Sklearn} package to extract features. For the \textit{PolitiFact} dataset, the dimensions of initial features of news articles, creators, and subjects are 3000, 3109, and 191 respectively. For the \textit{BuzzFeed} dataset, the parameter \textit{max\_features} for the news article nodes is set as 3000. 

\subsubsection{Comparison Methods}\label{sec:comparison_method}
We classify comparison methods into three categories: Graph neural network methods, Text classification methods, and Network embedding methods.

\noindent
\textbf{\textit{Graph neural network methods}}
\begin{itemize}[leftmargin=*]
	\item \textbf{\our}: {\our} is the proposed model.
	\item \textbf{$\our_{entropy}$}: We keep the active learning setting of {\our}, but query the candidates according to entropy. Here, we define that the closer the probability of this node being fake news to 0.5, the higher its entropy.
	\item \textbf{$\our_{random}$}: Here, we query the candidates for active learning randomly. 
	\item \textbf{HGAT-based classifier}: It is the classifier in the proposed {\our}. We test the performance without active learning setting. 
	\item \textbf{HAN} \cite{YYDHSH16}:  HAN employs node-level attention and semantic-level attention to capture the information from all meta-paths. In our experiments, we utilize two meta-paths (article-creator-article, article-subject-article) in HAN.
	\item \textbf{GAT} \cite{VCCRLB18}:  GAT is also an attention-based graph neural network for the node classification, but it is designed for homogeneous graphs. The News-HIN is treated as a homogeneous graph  (ignore the type information) when testing the model.
	\item \textbf{GCN}  \cite{TM17}: GCN is a semi-supervised methods for the node classification in homogeneous graphs. The News-HIN is treated as a homogeneous graph when testing it.
	
\end{itemize}
\noindent
\textbf{\textit{Text classification methods}}
\begin{itemize}[leftmargin=*]
	%	\vspace{-2pt}
	\item \textbf{SVM}: SVM is a classic supervised learning model. The feature vector used for building the SVM model is extracted merely based on the news article contents with TF-IDF. 
	\item \textbf{Text-CNN} \cite{kim2014convolutional}: Text-CNN is a text classification method based on convolutional neural network. It utilizes convolution filters of various sizes to capture key local features in news contents.
	%	\item \textbf{RLANS} \cite{li2018learning}:
	\item \textbf{LIWC} \cite{PBJB15}: LIWC stands for Linguistic Inquiry and Word Count, which is widely used to extract the lexicons falling into psycho-linguistic categories. It learns a feature vector from psychology and deception perspective.
\end{itemize}
\noindent
\textbf{\textit{Network embedding methods (NE)}}
\begin{itemize}[leftmargin=*]
	%	\vspace{-2pt}
	\item \textbf{Label Propagation (LP)} \cite{Zhu02learningfrom}: LP is merely based on the network structure. The prediction scores will be rounded and cast into labels.
	\item \textbf{DeepWalk} \cite{PAS14}: A random walk based embedding method, which is designed to deal with the homogeneous network. Based on the embedding results, we then train a logistic regression model to perform the classification of news articles.
	\item \textbf{LINE} \cite{TQWZYM15}: LINE optimizes the objective function that preserves the local and the global network structure simultaneously. We also learn a logistic regression model to conduct the classification based on the learned embeddings.
\end{itemize}

%Among these baseline methods, \textbf{SVM}, \textbf{LIWC}, \textbf{Text-CNN} use the text information of news article nodes only, while \textbf{DeepWalk}, \textbf{LINE} and \textbf{Label Propagation} learn from the network structure merely but ignore the text and node type information. Graph neural network methods can make use of textual contents and network structure simultaneously.

We have also noticed some recently appeared methods for fake news detection \cite{cui2019defend,SWL19,qian2018neural}, but did not compare them. The main consideration is the difference between the scenarios we face. In above works, they all utilize social context like user comments, but {\our} aims at detecting fake news in a relatively early stage with less labeled data. We won't utilize user comments about the news or large amount of training data, because when many users have started to discuss one fake news, the bad influence of fake news has spread.

%\vspace{-10pt}
%\subsection{Reproducibility}
\vspace{-5pt}
\subsection{Experimental Results with Analysis}

%-----------------------------------------------------------------------
\begin{table*}[t]
	\vspace{-25pt}
	\caption{Performance comparison of different methods. The training ratio is 20\%.}
	\label{tab:main_result_fix_train_ratio}
	\centering
	\vspace{-5pt}
	%\scriptsize
	\setlength{\tabcolsep}{3pt}
	{
		
		\begin{tabular}{lrcccccccc}
			\toprule
			\midrule
			\multicolumn{2}{l}{}&\multicolumn{4}{c}{\textit{PolitiFact}}&\multicolumn{4}{c}{\textit{BuzzFeed}}\\
			\cmidrule{3-10}
			& Methods	&Accuracy	& Precision	& Recall & F1 &Accuracy	& Precision	& Recall & F1\\
			\midrule
			\multirow{3}{*}{\rotatebox{90}{Text}} &SVM	&0.5432     &0.4975    &0.32     &0.3894  &0.5398     &0.6011    &0.5109     &0.5523  	\\
			&LIWC	&0.4544     &0.4415    &0.23     &0.3023   &0.6137    &0.6459    &0.5885     &0.6175	\\
			&Text-CNN	&0.5658     &0.5873    &0.2824     &0.3814	  	&0.6317     &0.6415    &0.6233    &0.6322\\
			\midrule
			\multirow{3}{*}{\rotatebox{90}{NE}} &Label Propagation	&0.5796    &0.7005    &0.1164     &0.1996 &0.5867    &0.6409    &0.223     &0.3309  	\\
			&DeepWalk	&0.5297     &0.4639    &0.2881    &0.4639  	&0.3721     &0.3083   &0.4322    &0.3599  	\\
			&LINE	&0.5012     &0.4109    &0.1215     &0.4109  &0.5899     &0.6123    &0.3057     &0.4077  	\\
			\midrule
			\multirow{7}{*}{\rotatebox{90}{GNNs}} &GAT	&0.5765     &0.7569    &0.0453     &0.0854 	&0.5885    &0.654    &0.3367     &0.4445  	\\
			&GCN	&0.5611     &\textbf{0.9688}    &0.0246     &0.048  &0.5671    &0.6331    &0.2674     &0.3816 	\\
			&HAN	&0.5867     &0.6802    &0.2062     &0.3165  &0.5917     &0.7163    &0.4677     &0.5659  	\\
			\cmidrule[2pt]{2-10}
			&HGAT-based classifier	&0.6154     &0.578    &0.424     &0.4893 	&0.7022     &0.6928   &0.6412     &0.666	\\
			&$\our_{random}$	&0.5724     &0.5152    &0.5515    &0.5328  &0.6843    &0.6439    &0.6123   &0.6277	\\
			&$\our_{entropy}$	&0.5601     &0.5022    &0.5581     &0.5286  &0.7161     &0.7088    &0.6503     &0.6783	\\
			&\our	&\textbf{0.6155}     &0.5661    &\textbf{0.5804}     &\textbf{0.5732} &\textbf{0.7351}     &\textbf{0.7211}    &\textbf{0.6909}     &\textbf{0.7057} 	\\
			\midrule
			\bottomrule
		\end{tabular}
		\vspace{-10pt}
	}
	
\end{table*}

%-----------------------------------------------------------------------
\begin{table}[h]
	%\vspace{-40pt}
	\caption{Adversarial Active Learning Performance of {\our} in \textit{PolitiFact}}
	\label{tab:main_result_AA_learning}
	\centering
	\vspace{-5pt}
	\scriptsize
	\setlength{\tabcolsep}{1.2pt}
	{
		\begin{tabular}{lrcccccccccc}
			\toprule
			\midrule
			\multicolumn{2}{l}{}&\multicolumn{8}{c}{Number of training nodes}\\
			\cmidrule{3-12}
			&Metrics	&1000	& 1200	& 1400	& 1600	&1800	&2000	&2200	& 2440	& 2600 & 2800 \\
			\midrule
			
			&Accuracy	&0.5658	&0.5878	&0.6049	&0.6053	&0.6013	&0.5984	&0.597	&0.597	&0.5955	&0.6155	\\
			&Precision &0.5142	&0.5246	&0.5218	&0.5245	&0.5135	&0.5115	&0.516	&0.5136	&0.5342	&0.5661 	\\
			&Recall	&0.3241	&0.4526	&0.4869	&0.5065	&0.5277	&0.5441	&0.5539	&0.5523	&0.5688	&0.5804	\\
			&F1	&0.3975	&0.4859	&0.5038	&0.5154	&0.5205	&0.5273	&0.5342	&0.5323	&0.5456	&0.5732	\\
			\midrule
			\bottomrule
		\end{tabular}
		\vspace{-5pt}
	}
	
\end{table}
\subsubsection{Assessing Impact of News-HIN}
In order to answer \textbf{Question 1}, we first present experiment results in Table~\ref{tab:main_result_fix_train_ratio} to compare AA-HGNN with three categories of methods introduced in Section~\ref{sec:comparison_method}. For text classification methods SVM, LIWC and Text-CNN which use the textual information of news article nodes to do classification, we see that Text-CNN \textgreater SVM \& LIWC in all metrics. This result shows that Text-CNN can better capture the important textual features in news contents by utilizing multiple convolution filters.
For Network embedding methods relying on graph structures, all of them achieve a poor recall. Recall is a pretty critical metric for fake news detection problem. A low recall means we omit lots of fake news so that they will cause bad social influence, which is unexpected. 
A News-HIN integrates all heterogeneous available data in the form of a graph structure. Intuitively, methods ({\our}, HAN) making full use of New-HIN as training data achieve better results. 
Through the comparison among GNNs methods, we verify that the heterogeneity of networks should be dealt with in a more effective way. If we simply treat a heterogeneous network as a homogeneous network by ignoring the type, then the results (reported by GAT, GCN) would be very disappointing. We continue to discuss performance concerning heterogeneity in the next section.
%-----------------------------------------------------------------
\begin{figure}[h]
	\vspace{-12pt}
	\centering
	\begin{minipage}[l]{0.7\columnwidth}
		\centering
		\includegraphics[width=\textwidth]{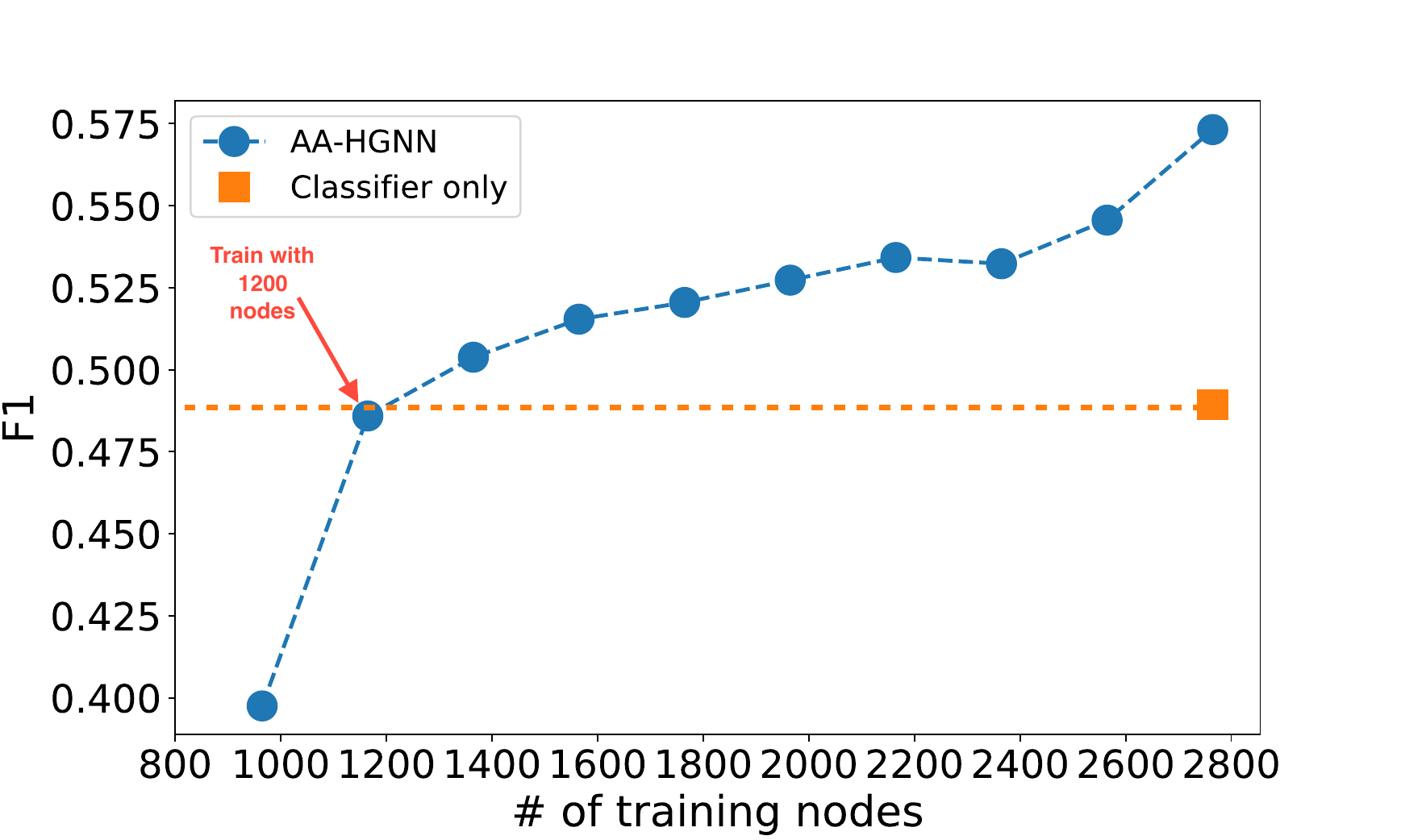}
	\end{minipage}
	%	\vspace{-10pt}
	\caption{The advantage in training with less labeled data.}\label{fig:train_less_data} 
	\vspace{-15pt}
\end{figure}

%-----------------------------------------------------------------
\begin{figure}[h]
	\vspace{-15pt}
	\centering
	\subfigure[F1]{ \label{fig:budget_f1}
		\begin{minipage}[l]{0.46\columnwidth}
			\centering
			\includegraphics[width=1.1\textwidth,height=3.2cm]{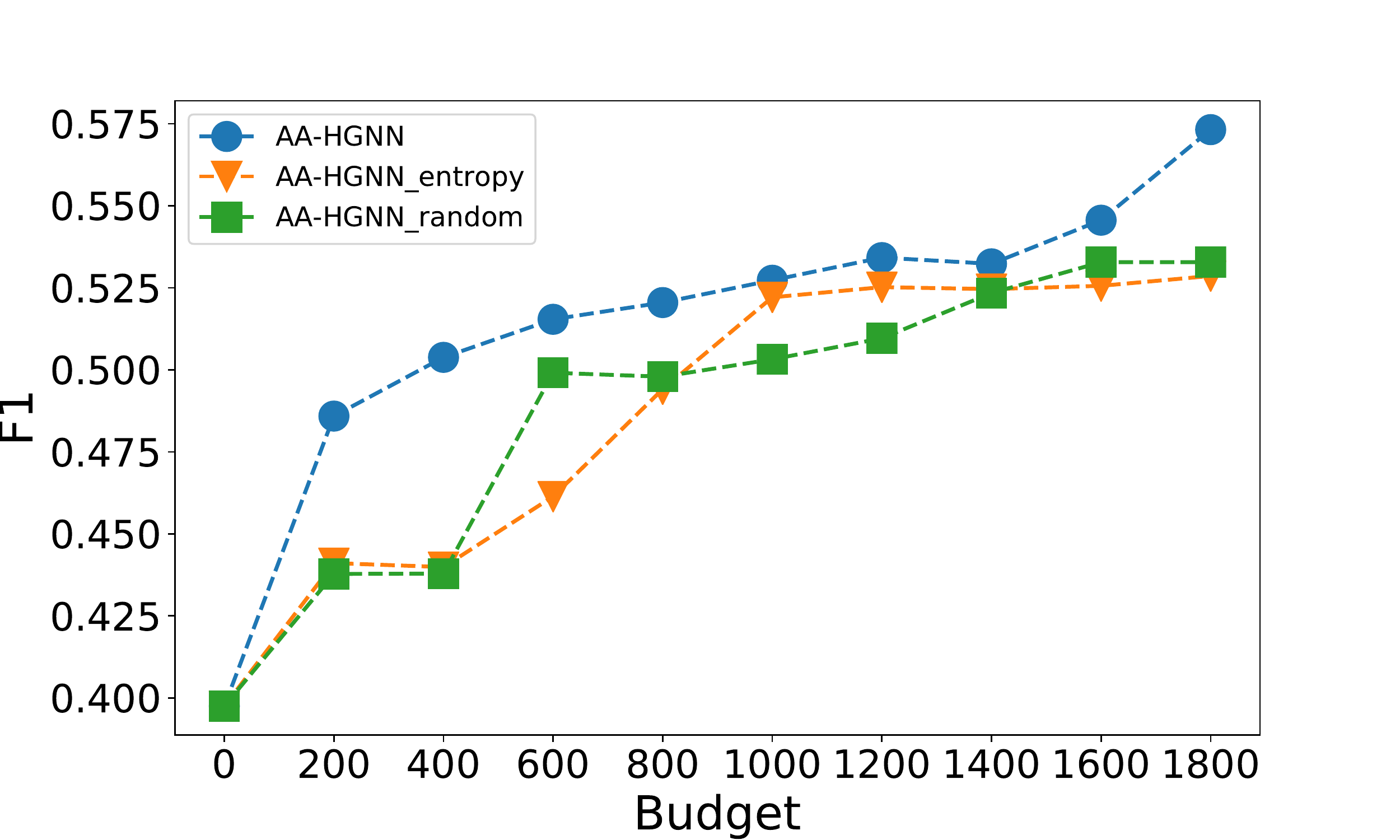}\vspace{3pt}
		\end{minipage}
	}
	\subfigure[Recall]{\label{fig:budget_recall}
		\begin{minipage}[l]{0.46\columnwidth}
			\centering
			\includegraphics[width=1.1\textwidth,height=3.2cm]{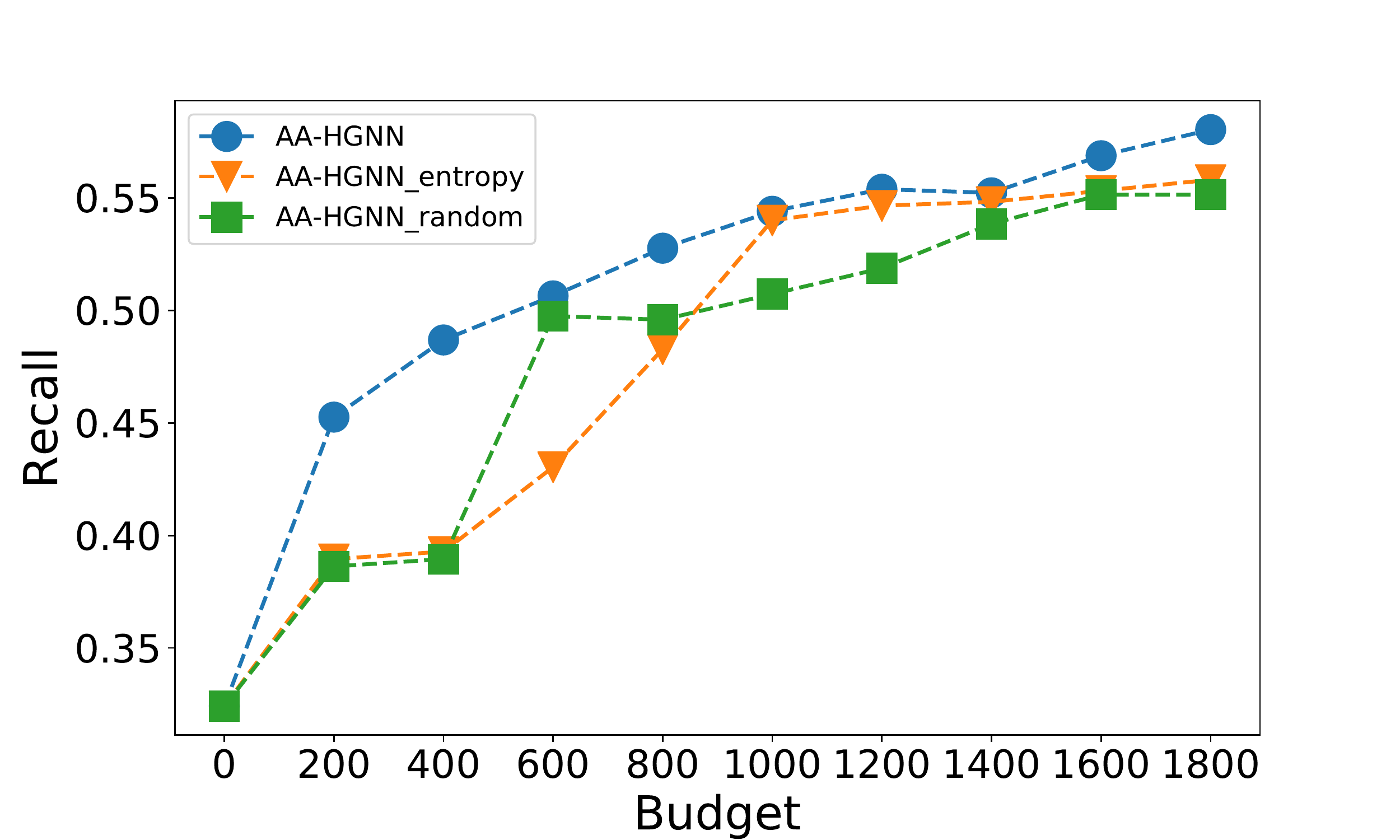}\vspace{3pt}
		\end{minipage}
	}
	\subfigure[Precision]{ \label{fig:budget_precision}
		\begin{minipage}[l]{0.46\columnwidth}
			\centering
			\includegraphics[width=1.1\textwidth,height=3.2cm]{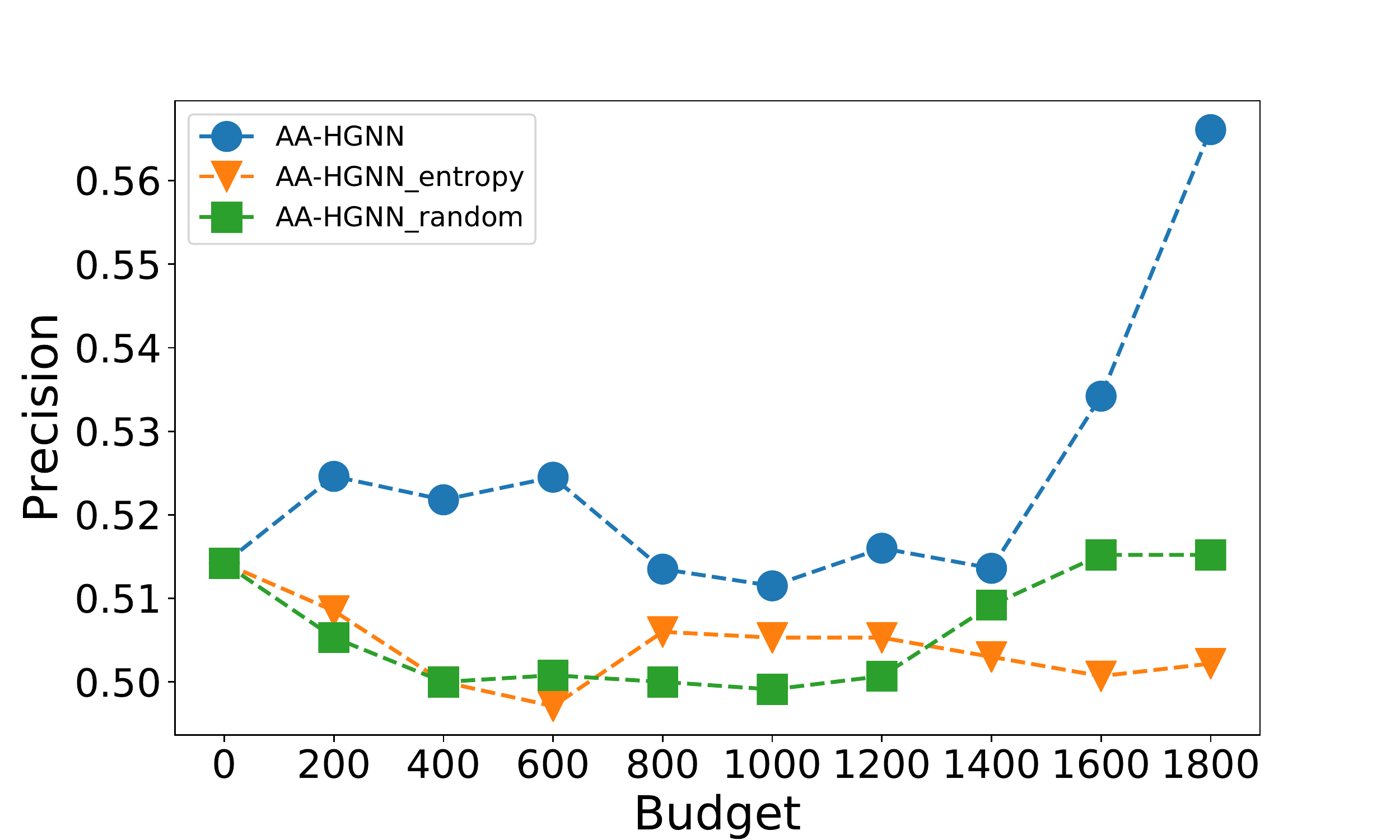}\vspace{3pt}
		\end{minipage}
	}
	\subfigure[Accuracy]{ \label{fig:budget_accuracy}
		\begin{minipage}[l]{0.46\columnwidth}
			\centering
			\includegraphics[width=1.1\textwidth,height=3.2cm]{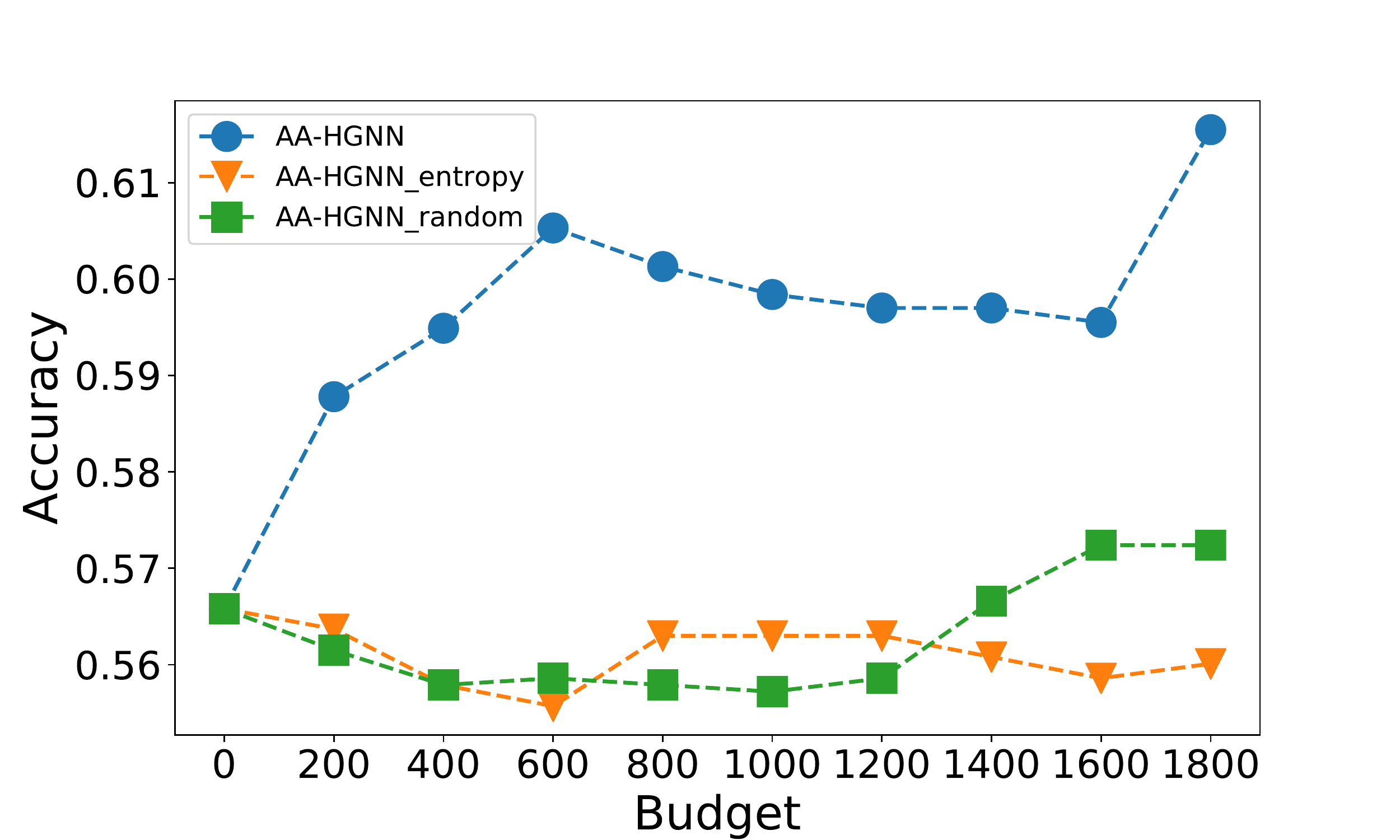}\vspace{3pt}
		\end{minipage}
	}
	%	\vspace{-15pt}
	\caption{Performance Analysis of Query Strategy in \textit{PolitiFact}}\label{fig:query_strategy_analysis}
	\vspace{-20pt}
\end{figure}

\subsubsection{Methods performance on Heterogeneous graph}
To answer \textbf{Question 2}, we further investigate the performance of different GNN methods besides AA-HGNN and its variants. As we utilize a heterogeneous network as source data, the heterogeneity should be handle in an effective manner. In Table~\ref{tab:main_result_fix_train_ratio}, we observe HGAT achieves the best accuracy, recall and F1. GAT and GCN get high precision but low recall. Particularly for the \textit{PolitiFact} dataset, GCN reach 0.9688 in precision but 0.0246 in recall. This result occurs because they prefer to classify a sample as real news based on News-HIN. They are not powerful methods in fake news problem because they were originally designed for homogeneous networks. Also as a method for heterogeneous graphs, HGAT-based classifier also shows an advantage over HAN. As the basic classifier, HGAT-based classifier can handle the heterogeneity of News-HIN well. 

\subsubsection{Active learning setting on scarce training data}
To answer \textbf{Question 3}, we draw Figure~\ref{fig:train_less_data} to compare the performance of HGAT-based classifier and AA-HGNN. The F1 score of the classifier shown in Figure~\ref{fig:train_less_data} is achieved with 2800 training nodes. In comparison, {\our} can outperform the classifier when being trained with 1200 labeled nodes. Besides, the score of {\our} applying the active learning setting significantly increased. When the number of training nodes is 2800, the performance of AA-HGNN increase nearly 9\% than the model without the active learning setting. From Table~\ref{tab:main_result_fix_train_ratio}, we can observe that {\our} has the apparent advantage when using 20\% training ratio, while other mehtods can not perform well due to the paucity of training data. Also, we see {\our} can reach satisfactory result although the training data is even more scarce in Table~\ref{tab:main_result_AA_learning}.

\subsubsection{Adversarial learning impacts on Active Learning}
In order to answer \textbf{Question 4}, we build two variants $\our_{entropy}$ and $\our_{random}$ to demonstrate the adversarial learning setting's efforts. These two varients provide different query strategies for active learning. Based on the results of comparative experiments in Figure~\ref{fig:query_strategy_analysis}, it is obvious that {\our} outperforms $\our_{entropy}$ and $\our_{random}$ in every query batch. The adversarial learning between the classifier and the selector indeed provides an effective query strategy for the active learning. The queried candidates are of high value for the classifier, so the performance of the classifier can be significantly improved. Besides, the adversarial learning-based query strategy can consistently provide high-value candidates, as the performance of selectors also improves in adversarial learning.

%-----------------------------------------------
\vspace{-5pt}
\section{Conclusion}\label{sec:conclusion}
\vspace{-5pt}
In this paper, we study the HIN-based fake news detection problem and propose a novel adversarial active learning-based graph neural network {\our} to solve it. {\our} employs a novel hierarchical attention mechanism to deal with the heterogeneity of News-HIN and learns textual and structural information simultaneously. An active learning framework is applied in {\our} to enhance the learning performance, especially when facing the paucity of labeled data. A selector is trained in an adversarial manner to query high-value candidates for the active learning setting. Experiments with real-world fake news data show that our model can outperform text-based models and other graph-based models when using less labeled data. Experiments also verify the effectiveness of adversarial learning-based query strategy, which consistently queries high-value candidates to improve the performance. As an adversarial active learning-based model, {\our} is ideal for detecting fake news in the early stages when lacking training data. Finally, due to the good generalizability of {\our}, it has the ability to be widely used in other node classification-related applications on heterogeneous graphs, where  there will be no obstacles to the transfer.

%-----------------------------------------------
\section{Acknowledgement}\label{sec:ack}

This work is partially supported by NSF through grant IIS-1763365 and by FSU.
%-----------------------------------------------
\balance
\bibliographystyle{plain}
\bibliography{reference}

\begin{thebibliography}{10}

\bibitem{aggarwal2014active}
Charu~C Aggarwal, Xiangnan Kong, Quanquan Gu, Jiawei Han, and S~Yu Philip.
\newblock Active learning: A survey.
\newblock In {\em Data Classification}, pages 599--634. Chapman and Hall/CRC,
  2014.

\bibitem{AG17}
H.~Allcott and M.~Gentzkow.
\newblock Social media and fake news in the 2016 election.
\newblock {\em Journal of Economic Perspectives}, 2017.

\bibitem{PAS14}
Rami Al-Rfou B.~Perozzi and Steven Skiena.
\newblock Deepwalk: Online learning of social representations.
\newblock In {\em KDD}, 2014.

\bibitem{bastos2019brexit}
Marco~T Bastos and Dan Mercea.
\newblock The brexit botnet and user-generated hyperpartisan news.
\newblock {\em Social Science Computer Review}, 37(1):38--54, 2019.

\bibitem{JWAY13}
Joan Bruna, Wojciech Zaremba, Arthur Szlam, and Yann LeCun.
\newblock Spectral networks and locally connected networks on graphs.
\newblock In {\em arXiv preprint arXiv:1312.6203}, 2013.

\bibitem{GPLJFA15}
Giovanni~Luca Ciampaglia, Prashant Shiralkar, Luis~M Rocha, Johan Bollen,
  Filippo Menczer, and Alessandro Flammini.
\newblock Computational fact checking from knowledge networks.
\newblock {\em PloS one}, 2015.

\bibitem{cui2019defend}
Limeng Cui, Kai Shu, Suhang Wang, Dongwon Lee, and Huan Liu.
\newblock defend: A system for explainable fake news detection.
\newblock In {\em Proceedings of the 28th ACM International Conference on
  Information and Knowledge Management}, pages 2961--2964. ACM, 2019.

\bibitem{dai2018adversarial}
Quanyu Dai, Qiang Li, Jian Tang, and Dan Wang.
\newblock Adversarial network embedding.
\newblock In {\em Thirty-Second AAAI Conference on Artificial Intelligence},
  2018.

\bibitem{dasgupta2005analysis}
Sanjoy Dasgupta, Adam~Tauman Kalai, and Claire Monteleoni.
\newblock Analysis of perceptron-based active learning.
\newblock In {\em International Conference on Computational Learning Theory},
  pages 249--263. Springer, 2005.

\bibitem{DGHH14}
Xin Dong, Evgeniy Gabrilovich, Geremy Heitz, Wilko Horn, NiLao, Kevin Murphy,
  Thomas Strohmann, Shaohua Sun, and Wei Zhang.
\newblock Knowledge vault: A web-scale approach to probabilistic knowledge
  fusion.
\newblock In {\em KDD}, 2014.

\bibitem{goodfellow2014generative}
Ian Goodfellow, Jean Pouget-Abadie, Mehdi Mirza, Bing Xu, David Warde-Farley,
  Sherjil Ozair, Aaron Courville, and Yoshua Bengio.
\newblock Generative adversarial nets.
\newblock In {\em Advances in neural information processing systems}, pages
  2672--2680, 2014.

\bibitem{MGF05}
Marco Gori, Gabriele Monfardini, and Franco Scarselli.
\newblock A new model for learning in graph domains.
\newblock In {\em IJCNN}, 2005.

\bibitem{GCZGL18}
Han Guo, Juan Cao, Yazi Zhang, Junbo Guo, and Jintao Li.
\newblock Rumor detection with hierarchical social attention network.
\newblock In {\em CIKM}, 2018.

\bibitem{gupta2013faking}
Aditi Gupta, Hemank Lamba, Ponnurangam Kumaraguru, and Anupam Joshi.
\newblock Faking sandy: characterizing and identifying fake images on twitter
  during hurricane sandy.
\newblock In {\em Proceedings of the 22nd international conference on World
  Wide Web}, pages 729--736. ACM, 2013.

\bibitem{SRMS18}
Shashank Gupta, Raghuveer Thirukovalluru, Manjira Sinha, and Sandya
  Mannarswamy.
\newblock Cimtdetect: A community infused matrix-tensor coupled factor- ization
  based method for fake news detection.
\newblock In {\em arXiv preprint arXiv:1809.05252}, 2018.

\bibitem{WRJ17}
William~L. Hamilton, Rex Ying, and Jure Leskovec.
\newblock Inductive representation learning on large graphs.
\newblock In {\em NIPS}, 2017.

\bibitem{hu2019adversarial}
Binbin Hu, Yuan Fang, and Chuan Shi.
\newblock Adversarial learning on heterogeneous information networks.
\newblock In {\em Proceedings of the 25th ACM SIGKDD International Conference
  on Knowledge Discovery \& Data Mining}, 2019.

\bibitem{JCZL16}
Zhiwei Jin, Juan Cao, Yongdong Zhang, and Jiebo Luo.
\newblock News verification by exploiting conflicting social viewpoints in
  microblogs.
\newblock In {\em AAAI}, 2016.

\bibitem{kim2014convolutional}
Yoon Kim.
\newblock Convolutional neural networks for sentence classification.
\newblock {\em arXiv preprint arXiv:1408.5882}, 2014.

\bibitem{TM17}
Thomas~N. Kipf and Max Welling.
\newblock Semi-supervised classi cation with graph convolutional networks.
\newblock In {\em ICLR}, 2017.

\bibitem{li2018learning}
Yan Li and Jieping Ye.
\newblock Learning adversarial networks for semi-supervised text classification
  via policy gradient.
\newblock In {\em Proceedings of the 24th ACM SIGKDD International Conference
  on Knowledge Discovery \& Data Mining}, pages 1715--1723. ACM, 2018.

\bibitem{YDMR16}
Yujia Li, Daniel Tarlow, Marc Brockschmidt, and Richard Zemel.
\newblock Gated graph sequence neural networks.
\newblock In {\em ICLR}, 2016.

\bibitem{mendoza2010twitter}
Marcelo Mendoza, Barbara Poblete, and Carlos Castillo.
\newblock Twitter under crisis: Can we trust what we rt?
\newblock In {\em Proceedings of the first workshop on social media analytics},
  pages 71--79. ACM, 2010.

\bibitem{PBJB15}
J.~Pennebaker, R.~Boyd, K.~Jordan, and K.~Blackburn.
\newblock The development and psychometric properties of liwc.
\newblock {\em Technical Report}, 2015.

\bibitem{VBAR17}
Veronica Perez-Rosas, Bennett Kleinberg, Alexandra Lefevre, and Rada Mihalcea.
\newblock Automatic detection of fake news.
\newblock In {\em arXiv preprint arXiv:1708.07104}, 2017.

\bibitem{MJKJB17}
Martin Potthast, Johannes Kiesel, Kevin Reinartz, Janek Bevendor, and Benno
  Stein.
\newblock A stylometric inquiry into hyperpartisan and fake news.
\newblock In {\em arXiv preprint arXiv:1702.05638}, 2017.

\bibitem{qian2018neural}
Feng Qian, Chengyue Gong, Karishma Sharma, and Yan Liu.
\newblock Neural user response generator: Fake news detection with collective
  user intelligence.
\newblock In {\em IJCAI}, pages 3834--3840, 2018.

\bibitem{Rapoza17}
K.~Rapoza.
\newblock Can 'fake news' impact the stock market?
\newblock {\em Forbes News}, 2017.

\bibitem{ren2019activeiter}
Yuxiang Ren, Charu Aggarwal, and Jiawei Zhang.
\newblock Activeiter: Meta diagram based active learning in social networks
  alignment.
\newblock {\em IEEE Transactions on Knowledge and Data Engineering}, 2019.

\bibitem{ren2019meta}
Yuxiang Ren, Charu~C Aggarwal, and Jiawei Zhang.
\newblock Meta diagram based active social networks alignment.
\newblock In {\em 2019 IEEE 35th International Conference on Data Engineering
  (ICDE)}, pages 1690--1693. IEEE, 2019.

\bibitem{ren2019heterogeneous}
Yuxiang Ren, Bo~Liu, Chao Huang, Peng Dai, Liefeng Bo, and Jiawei Zhang.
\newblock Heterogeneous deep graph infomax.
\newblock {\em arXiv preprint arXiv:1911.08538}, 2019.

\bibitem{ren2020hgat}
Yuxiang Ren and Jiawei Zhang.
\newblock Hgat: hierarchical graph attention network for fake news detection.
\newblock {\em arXiv preprint arXiv:2002.04397}, 2020.

\bibitem{rubens2015active}
Neil Rubens, Mehdi Elahi, Masashi Sugiyama, and Dain Kaplan.
\newblock Active learning in recommender systems.
\newblock In {\em Recommender systems handbook}, pages 809--846. Springer,
  2015.

\bibitem{VRTL15}
Victoria~L Rubin and Tatiana Lukoianova.
\newblock Truth and deception at the rhetorical structure level.
\newblock {\em Journal of the Association for Information Science and
  Technology}, 2015.

\bibitem{NSY18}
Natali Ruchansky, Sungyong Seo, and Yan Liu.
\newblock Csi: A hybrid deep model for fake news detection.
\newblock In {\em CIKM}, 2017.

\bibitem{FMAMG09}
Franco Scarselli, Marco Gori, Ah~Chung Tsoi, Markus Hagenbuchner, and Gabriele
  Monfardini.
\newblock The graph neural network model.
\newblock {\em IEEE Transactions on Neural Networks 20, 1}, 2009.

\bibitem{serratosa2015interactive}
Francesc Serratosa and Xavier Cort{\'e}s.
\newblock Interactive graph-matching using active query strategies.
\newblock {\em Pattern Recognition}, 48(4):1364--1373, 2015.

\bibitem{BT14}
Baoxu Shi and Tim Weninger.
\newblock Discriminative predicate path mining for fact checking in knowledge
  graphs.
\newblock {\em Knowledge-Based Systems}, 2014.

\bibitem{SLZSY17}
C.~Shi, Y.~Li, J.~Zhang, Y.~Sun, and P.~Yu.
\newblock A survey of heterogeneous information network analysis.
\newblock {\em TKDE}, 2017.

\bibitem{SWL19}
Kai Shu, Suhang Wang, and Huan Liu.
\newblock Beyond news contents: The role of social context for fake news
  detection.
\newblock In {\em WSDM}, 2019.

\bibitem{shu2019beyond}
Kai Shu, Suhang Wang, and Huan Liu.
\newblock Beyond news contents: The role of social context for fake news
  detection.
\newblock In {\em Proceedings of the Twelfth ACM International Conference on
  Web Search and Data Mining}, pages 312--320, 2019.

\bibitem{SH12}
Y.~Sun and J.~Han.
\newblock Mining heterogeneous information networks: a structural analysis
  approach.
\newblock {\em KDD Explorations}, 2012.

\bibitem{TQWZYM15}
Jian Tang, Meng Qu, Mingzhe Wang, Ming Zhang, Jun Yan, and Qiaozhu Mei.
\newblock Line: Large-scale information network embedding.
\newblock In {\em WWW}, 2015.

\bibitem{VCCRLB18}
P.~Velickovic, G.~Cucurull, A.~Casanova, A.~Romero, P.~Lio, and Y.~Bengio.
\newblock Graph attention networks.
\newblock In {\em ICLR}, 2018.

\bibitem{vosoughi2018spread}
Soroush Vosoughi, Deb Roy, and Sinan Aral.
\newblock The spread of true and false news online.
\newblock {\em Science}, 359(6380):1146--1151, 2018.

\bibitem{wang2017irgan}
Jun Wang, Lantao Yu, Weinan Zhang, Yu~Gong, Yinghui Xu, Benyou Wang, Peng
  Zhang, and Dell Zhang.
\newblock Irgan: A minimax game for unifying generative and discriminative
  information retrieval models.
\newblock In {\em Proceedings of the 40th International ACM SIGIR conference on
  Research and Development in Information Retrieval}, pages 515--524. ACM,
  2017.

\bibitem{wang2016cost}
Keze Wang, Dongyu Zhang, Ya~Li, Ruimao Zhang, and Liang Lin.
\newblock Cost-effective active learning for deep image classification.
\newblock {\em IEEE Transactions on Circuits and Systems for Video Technology},
  27(12):2591--2600, 2016.

\bibitem{WJSWCYY19}
Xiao Wang, Houye Ji, Chuan Shi, Bai Wang, Peng Cui, P.~Yu, and Yanfang Ye.
\newblock Heterogeneous graph attention network.
\newblock In {\em WWW}, 2019.

\bibitem{XHLJ10}
K.~Xu, W.~Hu, J.~Leskovec, and S.~Jegelka.
\newblock How powerful are graph neural networks?
\newblock In {\em ICLR}, 2019.

\bibitem{YYDHSH16}
Z.~Yang, D.~Yang, C.~Dyer, X.~He, A.~Smola, and E.~Hovy.
\newblock Hierarchical attention networks for document classification.
\newblock In {\em NAACL}, 2016.

\bibitem{ZCFG18}
Jiawei Zhang, Limeng Cui, Yanjie Fu, and Fisher~B Gouza.
\newblock Fake news detection with deep diffusive network model.
\newblock In {\em arXiv preprint arXiv:1805.08751}, 2018.

\bibitem{ZZ18}
Xinyi Zhou and Reza Zafarani.
\newblock Fakenews: A survey of research, detection methods, and opportunities.
\newblock In {\em arXiv preprint arXiv:1812.00315}, 2018.

\bibitem{Zhu02learningfrom}
Xiaojin Zhu and Zoubin Ghahramani.
\newblock Learning from labeled and unlabeled data with label propagation.
\newblock Technical report, 2002.

\end{thebibliography}
%-----------------------------------------------
%\newpage
%\input{sec_appendix}
\end{document}